\documentclass[twocolumn,notitlepage,superscriptaddress, nofootinbib,nobibnotes, aps,prd,10pt]{revtex4-1}
\usepackage{amsmath,amssymb, fp}  
\usepackage{xcolor}  
\usepackage{soul} 
\definecolor{linkcolor}{RGB}{150,50,60}  
\usepackage[	colorlinks=true,    	
			pdfstartview=FitV,
			linkcolor= linkcolor,
			citecolor= linkcolor,
			urlcolor= linkcolor,
			linktoc=page,
			hyperindex=true,
			hyperfigures=true,  
			breaklinks=true]
			{hyperref}
\usepackage{dblfloatfix}    
\usepackage{balance} 
\usepackage{lastpage}  
\usepackage{mathtools}
\usepackage{epsfig,rotating,pifont}
\usepackage{graphicx}
\usepackage[caption=false]{subfig}

\usepackage{ragged2e} 
\usepackage{etoolbox}
\usepackage{changepage}   
\usepackage{pgfplots}
\usetikzlibrary{pgfplots.groupplots}
\pgfplotsset{compat=1.3}
\usepackage{tikz}  
\usetikzlibrary{arrows,shapes,positioning}
\usetikzlibrary{decorations.markings,decorations.pathmorphing,decorations.pathreplacing,decorations.text}
\usetikzlibrary{arrows,calc,patterns,shapes.geometric}
\usepackage{fancybox}
\usepackage{verbatim}
\usepackage{multirow}
\usepackage{titlesec}

\def\beq{\begin{eqnarray}}
\def\eeq{\end{eqnarray}}

\def\a{\alpha}

\def\eps{\epsilon}

\def\la{\langle }
\def\ra{\rangle }

\def\lb{\label}

\def\M{\mathcal{M}}

\def\B{\mathcal{B}}
\def\O{\mathcal{O}}

\def\DM{\partial\mathcal{M}}

\def\cca{\mathfrak{a}}

\newcommand{\Tr}{\,\mathrm{Tr}\,}            

\newcommand{\be}{\begin{equation}}
\newcommand{\ee}{\end{equation}}
\newcommand{\bea}{\begin{eqnarray}}
\newcommand{\eea}{\end{eqnarray}}
\newcommand{\bg}{\begin{gather}}

\newcommand{\bseq}{\begin{subequations}}
\newcommand{\eseq}{\end{subequations}}

\newcommand{\ket}[1]{| #1 \rangle}

\def\tr{\hbox{tr}}

\def\be{\begin{eqnarray}}
\def\ee{\end{eqnarray}}
\def\lb{\label}

\def\emi{{\scalebox{0.6}{{\rm EMI}}}}
\def\bemi{{\scalebox{0.6}{\rm bEMI}}}
\def\LFT{{\scalebox{0.6}{\rm LFT}}}
\def\LFT{{{z=2}}}

\definecolor{Green}{RGB}{147,162,153}
\definecolor{Green2}{RGB}{26,148,49}
\definecolor{BrownL}{RGB}{173,143,103}
\definecolor{Red}{RGB}{210,83,60}
\definecolor{BrownD}{RGB}{114,96,86}
\definecolor{GreyD}{RGB}{76,90,106}
\definecolor{GreyB}{RGB}{128,141,160}
\definecolor{Maroon}{RGB}{121,70,61}
\definecolor{Blue}{RGB}{148,184,210}
\definecolor{Blue2}{RGB}{108,144,170}
\definecolor{Blue3}{RGB}{42, 107, 172}
\definecolor{BB}{RGB}{128,184,220}

\newsavebox\foobox
\begin{document}

\title{Relating bulk to boundary entanglement}  

\author{Cl\'ement Berthiere}  
\email{clement.berthiere@pku.edu.cn}
\address{Department of Physics, Peking University, Beijing 100871, China}  
\author{William Witczak-Krempa}
\email{w.witczak-krempa@umontreal.ca}
\address{D\'epartement de Physique, Universit\'e de Montr\'eal, Montr\'eal, Qu\'ebec, H3C 3J7, Canada}
\address{Centre de Recherches Math\'ematiques, Universit\'e de Montr\'eal; P.O. Box 6128, Centre-ville Station; Montr\'eal (Qu\'ebec), H3C 3J7, Canada}
\address{Regroupement Qu\'eb\'ecois sur les Mat\'eriaux de Pointe (RQMP)}

\date{\today} 
 
\begin{abstract}\vspace{4pt}
\begin{center}\textbf{\abstractname}\end{center}\vspace{-5pt} 
Quantum many-body systems have a rich structure in the presence of boundaries. We study the groundstates of conformal field theories (CFTs) and Lifshitz field theories in the presence of a boundary through the lens of the entanglement entropy.  
For a family of theories in general dimensions, we relate the universal terms in the entanglement entropy of the bulk theory with  
the corresponding terms for the theory with a boundary. This relation imposes a condition on certain boundary central charges. 
For example, in $2+1$ dimensions, we show that the corner-induced 
logarithmic terms of free CFTs and certain Lifshitz theories are simply related to those that arise when the corner touches the boundary.
We test our findings on the lattice, including a numerical implementation of Neumann boundary conditions.
We also propose an ansatz, the boundary Extensive Mutual Information model, for a CFT with a boundary whose entanglement entropy is purely geometrical. This model shows the same bulk-boundary connection as Dirac fermions and certain supersymmetric CFTs that have a holographic dual.
Finally, we discuss how our results can be generalized to all dimensions
as well as to massive quantum field theories.
\end{abstract}

\maketitle  

\makeatletter
\def\l@subsubsection#1#2{}
\makeatother
\tableofcontents

\section{Introduction} \label{sec:intro}       
Quantum many-body systems are often studied in infinite space or on spaces without boundaries, like tori and spheres, in order to simplify the
analysis. However, introducing a boundary is not only more realistic, but it can reveal novel phenomena. For instance, gapped topological phases like quantum Hall states often have protected boundary modes \cite{wen2007}.  
In fact, such topological boundary modes can often only exist at a boundary of a higher dimensional system. In the gapless realm that
  will be the focus of this work, boundaries can give rise to novel surface critical behaviors. Generally, many distinct
  boundary universality classes are possible for a given bulk one, which leads to new critical exponents that are absent in a bulk treatment, see e.g.\ \cite{Diehl:1996kd}.   

There has been a recent effort to understand the quantum entanglement properties of critical systems in the presence of a  
boundary, see for instance Refs.~\cite{Calabrese:2004eu,Calabrese:2009qy,2006PhRvL..96j0603L, 2009JPhA...42X4009A,Herzog:2015ioa,Fursaev:2016inw,Casini:2016fgb,Zhou:2016ykv,Chen:2016kjp,Chen:2017txi},     
which provides a new viewpoint compared to the study of correlation functions of local operators. This is partly   
  motivated by the success of entanglement measures in bulk systems. One example is the construction of a renormalization
  group monotone for relativistic theories in 3$d$ (where $d$ stands for the spacetime dimension)
  using the entanglement entropy for certain spatial bipartitions, i.e.\ the $F$-theorem \cite{sinha2010,sinha2011,Casini:2012ei}.
  We recall that the entanglement entropy associated with a pure state $|\psi \rangle$ and a subregion $A$ of the full space $A\cup B$ is defined as $S(A)=-\mbox{tr}(\rho_A\log \rho_A)$,
  where the reduced density matrix is $\rho_A=\mbox{tr}_B|\psi\rangle\langle\psi|$. 
  An extension of this work to relativistic systems with boundaries results in a new proof of the 
  $g$-theorem in 2$d$ \cite{Casini:2016fgb}, 
  and its generalization to higher dimensions \cite{Casini:2018nym}.
However, the entanglement structure and its dependence on boundary conditions remains largely unknown, the more so for non-relativistic theories.

In this work, we study the entanglement entropy (and its R\'enyi generalizations) in groundstates of
  gapless Hamiltonians in the presence of boundaries. An important role will be played by entangling surfaces
  that intersect the physical boundary. These lead to a new type of corner term that is distinct from the corner terms that have been
  extensively studied in the bulk. The entanglement entropy of such \emph{boundary corners} 
  has been studied for non-interacting CFTs \cite{Fursaev:2016inw, Berthiere:2016ott,Berthiere:2018ouo},
  certain interacting large-$N$ superconformal gauge theories  
  via the AdS$_{d+1}$/bCFT$_d$ correspondence \cite{Takayanagi:2011zk,Fujita:2011fp,FarajiAstaneh:2017hqv,Seminara:2017hhh,Seminara:2018pmr},
  and a special class of Lifshitz theories \cite{Fradkin:2006mb}. 
  For non-interacting CFTs we find that the boundary corner functions are directly related to the bulk  
  corner function via simple relations. We successfully verify our predictions numerically for the relativistic scalar on the lattice,
  which requires a numerical implementation of Neumann boundary conditions. For scalar and Dirac CFTs, we show that the boundary corner function
  can be used to extract certain boundary central charges.

  Our paper is organized as follows. After the Introduction, Section~\ref{sec:duality} introduces the relation between the entanglement entropy
    of bulk subregions to that of subregions in a theory with a physical boundary. In Section~\ref{sec:CFT}, we study the bulk-boundary relation for
    regions with corners in (boundary) CFTs, with a focus on free scalars and Dirac fermions. A numerical check on the lattice is presented for the scalar.
    In Section~\ref{sec:bEMI}, we propose an ansatz in general dimensions, the boundary Extensive Mutual Information model,
    for a CFT with a boundary whose entanglement entropy is purely geometrical. In three spacetime dimensions, we obtain the boundary corner function analytically, which gives
    a certain anomaly coefficient for the theory. In Section~\ref{sec:lif}, we study the entanglement properties of a gapless non-interacting Lifshitz theory.
    Using the heat kernel method, we obtain the boundary corner function for both Dirichlet and Neumann boundary conditions, and find that these have the
    same qualitative features as the relativistic scalar. In Section~\ref{sec:massive}, we discuss the extension of our results to massive quantum field theories, focusing on the relativistic scalar. We conclude in Section~\ref{sec:conclusion} with a summary of our main results, as well as an outlook on future research topics.
    Four appendices complete the paper: Appendix~\ref{apdxA} deals with central charges, Appendix~\ref{apdxB} discusses the entanglement entropy of cylindrical regions
    in $4d$ spacetimes for the relativistic scalar, Appendix~\ref{apdxC} shows our implementation of boundary conditions for the discretized scalar field (Dirichlet
    and Neumann), and Appendix~\ref{apdxD} recalls the high precision ansatz for the scalar bulk corner function.

\section{Relating bulk to boundary entanglement}  \label{sec:duality}
\subsection{$(1+1)$--dimensional systems}

For one--dimensional quantum systems of infinite length described by conformal theories, the $n$--R\'enyi entropy, $S_n(A)=\log(\mbox{tr}\rho_A^n)/(1-n)$,
of an interval of length $\ell$
takes the form \cite{Calabrese:2004eu,Calabrese:2009qy}
\be
S_n(\ell) = \frac{c}{6}\left(1+\frac{1}{n}\right)\log\frac{\ell}{\eps}+ 2c_n^0\,,\lb{EE2}
\ee
where $c$ is the central charge of the CFT, $\eps$ is a UV cut-off and $c_n^0$ is a non-universal constant. If the system is not infinite but has a boundary, say it is the semi-infinite line $[0,\infty[$, the R\'enyi entropies of a finite interval adjacent to the boundary $[0,\ell]$ are now given by \cite{Calabrese:2004eu,Calabrese:2009qy}
\be
S_n^{(\mathcal B)}(\ell) = \frac{c}{12}\left(1+\frac{1}{n}\right)\log\frac{2\ell}{\eps}+\log g_{\mathcal B} + c_n^0\,,\lb{bEE2}
\ee
where $\mathcal B$ is the boundary condition imposed at the origin, $c_n^0$ is the same \cite{2006PhRvA..74e0305Z} non-universal constant as in \eqref{EE2}, and $\log g_{\mathcal B}$ is the boundary entropy, first discussed by Affleck and Ludwig \cite{Affleck:1991tk} (see also \cite{2006PhRvL..96j0603L,2009JPhA...42X4009A}).

Looking at expressions \eqref{EE2} and \eqref{bEE2}, one immediately notices that the R\'enyi entropies for $2d$ CFTs and bCFTs satisfy
\be
S_n(2\ell)=2S_n^{(\mathcal B)}(\ell)\,,\lb{2d}
\ee 
at the leading order in $\eps$. 
Indeed, the logarithmically divergent part of the entropy of an interval in the presence of a boundary can be obtained from the entropy of the union of that interval with its mirror image (with respect to the boundary) in an infinite system, i.e. by the formula \eqref{2d} for an interval connected to the boundary.
In $2d$ bCFTs, the dependence of the $n$--R\'enyi entropy on the boundary conditions appears in the subleading terms to the logarithmic divergence, namely in the boundary entropy $\log g_\B$. Similarly, for $d$--dimensional CFTs, the presence of a boundary affects the terms subleading to the area law. This means that the analog of formula \eqref{2d} is valid at the area law level in higher dimensions, but does not necessarily hold for subleading terms, which are the
  interesting ones as they contain universal information. In this work,
  we shall show that such a relation between the universal part of the bulk and boundary entanglement entropies does exist in general dimensions.
  Our results cover not only free CFTs but also certain interacting ones, as well as Lifshitz theories.   

\pagebreak
  \subsection{Free CFTs in general dimensions}\lb{hkernel}
For free theories, the $n$--R\'enyi entropy may be computed using the heat kernel (or Green function) method together with the replica trick. Essentially, one has to compute the trace of the heat kernel on a manifold with a conical singularity along the entangling surface. Let us take the free scalar field as an example. For a base manifold that is the half-space in $\mathbb{R}^d$, we may impose either Dirichlet or Neumann BCs on the boundary (conformal BCs). 
The (scalar) heat kernel is then the sum\footnote{In one spatial dimension, the `uniform' term is the well-known solution of the heat equation on $\mathbb{R}$ with initial condition $K(0,x,x')=\delta(x-x')$, i.e. \mbox{$K(s,x,x') = \frac{1}{\sqrt{4\pi s}}e^{-\frac{1}{4s}(x-x')^2}$}, while the `reflected' term is the mirror image through the boundary at, e.g., $x=0$, that is $K^*(s,x,x')=K(s,x,-x')$. Also, $\tr\,K$ is the trace of the heat kernel over the manifold $\M$, $\tr\,K=\int_\M dx\, K(s,x,x)$.} of a `uniform' term, which equals the heat kernel $K$ on $\mathbb{R}^d$ (without boundary), and a `reflected' term $K^*$. The reflected term satisfies the heat equation, with boundary data canceling that of the uniform term. For Neumann $(+)$ and Dirichlet $(-)$ BCs, one has $K_{N/D}=K \pm K^*$. Taking the trace of these heat kernels one gets $\tr\, K=\widetilde{\tr} (K_N + K_D)$, where $\tr$ stands for the trace over $\mathbb{R}^d$ and $\widetilde{\tr}$ for the trace over the half-space only. Thus, considering the entropy of a scalar field for an arbitrary
subregion $A$ of $\mathbb{R}^d$ symmetric with respect to some hyperplane, one may obtain the entropy of $A$ as the sum of the Neumann and Dirichlet entanglement entropies of the two mirror subregions with a boundary being the hyperplane of symmetry of $A$. In $1+1$ dimensions, this reasoning leads to \eqref{2d}
at leading order in $\ell/\epsilon$ for free CFTs, independently of the boundary conditions.
As was discussed, this holds for general CFTs in 2$d$.
These considerations, along with new ones that we shall present in this work, motivate the following conjecture relating bulk and boundary entanglement in $d\ge2$.

\subsection{Bulk-boundary relation}   

Consider some arbitrary co-dimension 1 spatial region (not necessarily connected) in $\mathbb{R}^{1,d-1}$ which is symmetric with respect to a co-dimension 2 plane. In other words, this region is the union of two mirror symmetric regions $A$ and $A'$, as for example 
shown in Fig.\,\hyperref[fig1]{\ref{fig1}}. 
Then, for certain bQFTs, we conjecture that there exist some boundary conditions $\B$ and $\B'$ that may be imposed on the plane of symmetry
(physical boundary) such that the following relation between R\'enyi entropies holds  
\be
S_n(A\cup A') = S_n^{(\B)}(A) + S_n^{(\B')}(A')\,, \lb{dual} 
\ee 
where $S_n(A\cup A')$ is the $n$--R\'enyi entropy for the whole region \mbox{$A\cup A'$} in the spacetime without boundary, while $S_n^{(\B)}(A)$ is the $n$--R\'enyi entropy for the region $A$ with boundary condition $\B$ imposed on $\DM$, and similarly for $S_n^{(\B')}(A')$. One may think that \eqref{dual} strangely resembles the subadditivity property of an extensive configuration. However, it is not so because we compute entropies for different theories. 

A particular case of \eqref{dual} is given when the boundary conditions coincide, $\B=\B'$:
  \begin{align}
    S_n(A\cup A') = 2S_n^{(\B)}(A) \,, \lb{dual2} 
  \end{align}
  which can be seen as a generalization of \eqref{2d}. As we shall see, this form of the bulk-boundary entanglement relation
  will be realized for Dirac fermions, holographic CFTs, and the so-called (boundary) Extensive Mutual Information Model. 
\begin{figure}[h]
\centering
\includegraphics[scale=1]{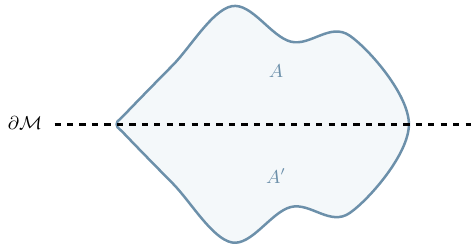}
\vspace{-3pt}
\caption{(b)CFT$_3$ on the (half-) plane. The region $A$ and its mirror image $A'$ with respect to the boundary $\DM$ (dashed line) are shown in blue.}
\lb{fig1}
\end{figure}

For $2d$ bCFTs, our relation \eqref{dual} would imply that \mbox{$g_{\B'}=g_{\B}^{-1}$} for certain pairs of boundary conditions $\B,\,\B'$. This is actually the case for the XX chain and free fermions with open boundary conditions for which $g_\B=1$ \cite{Affleck:1991tk,Fagotti:2010cc}. This condition on the
  boundary entropy can be seen as necessary for the bulk-boundary relation to hold beyond the leading logarithmic term. In higher dimensions,
  since the leading term in the R\'enyi entropy is the area law, we expect that the bulk-boundary relation implies a relation for a higher dimensional
  analogue of the boundary entropy. Let us consider the case of spacetime dimension $d=3$, which will be the focus of the present work.
  We consider our region $A$ to be a half-disk attached to the physical boundary $\DM$. Then its mirror image is also a half-disk, and
  $A\cup A'$ is a full disk, as illustrated in Fig.\,\hyperref[fig4]{\ref{fig4}}. The left hand side of \eqref{dual} for the groundstate of a CFT is then ($n=1$):
  \begin{align}
    S_1(A\cup A')= B\frac{2\pi R}{\epsilon} - F\,,\lb{Fdisk}
  \end{align}
  where $R$ is the radius of the disk, and the universal $R$-independent contribution features the RG monotone in $d=3$, $F$.
  In contrast, the right hand side of the relation \eqref{dual} will be built from the half-disk entropy
  \begin{align}
    S_1^{(\B)}(A) = B\frac{\pi R}{\epsilon} - s_{\rm log}^{(\B)}\log(R/\epsilon)+\cdots\,,
  \end{align}
  where we have omitted subleading terms in $R/\epsilon$. The logarithmic divergence comes from the two corners generated by the intersection of the entangling surface
  and the physical boundary. It was argued \cite{Fursaev:2016inw} that $s_{\rm log}^{(\B)}$ is proportional to the boundary central charge $\cca^\B$ that appears in the trace of the stress tensor
  as a consequence of the conformal anomaly.  
  We see that in order for the bulk-boundary entanglement relation \eqref{dual} at $n=1$ to hold, the logarithms must cancel, implying:
  \begin{align}
    \cca^\B + \cca^{\B'} = 0 \,.\lb{anomE}
  \end{align}
  For example, in the case of a free scalar field, the central charges for Dirichlet and Neumann boundary conditions have opposite sign, which is a necessary condition for the relation. If we are dealing with the relation for a single boundary condition $\B=\B'$, \eqref{dual2},
  this implies that the boundary central charge must vanish, $\cca^\B=0$.
  This will indeed be the case for Dirac fermions, certain holographic CFTs (with $\alpha=\pi/2$, see below), and the Extensive Mutual Information Model.
  It would be of interest to find which bCFTs obey the relation \eqref{anomE}, and the much stronger condition \eqref{dual}. One useful avenue would be to numerically
    investigate the quantum critical transverse field Ising model in two spatial dimensions along the lines of \cite{PhysRevLett.110.135702}. In any case, our conjectured relation \eqref{dual} provides
  a useful starting point to compare the bulk and boundary entanglement entropies of QFTs.

\section{CFTs in $2+1$ dimensions} \label{sec:CFT}

In two spatial dimensions, there are many ways to partition a domain. In this paper, we mainly study two different kind of regions that contain corners, and which produce a logarithmic correction to the area law in the entanglement entropy,  
\be
S=B \frac{\ell}{\eps}-s_{\rm log}(\theta)\log\frac{\ell}{\eps}+\cdots\,,
\ee
with a certain corner function $s_{\rm log}(\theta)$ as the cut-off independent coefficient of the logarithmic term. The two corner geometries of interest are depicted in Fig.\,\hyperref[fig2]{\ref{fig2}}. They may be classified according to whether they touch the boundary of the space (boundary corner), or not (bulk corner). 
%
\begin{figure}[h]
\vspace{-20pt}
\centering
\subfloat[]{
\includegraphics[scale=1]{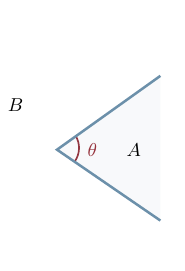}
}
\hspace{0.9cm}
\subfloat[]{
\includegraphics[scale=1]{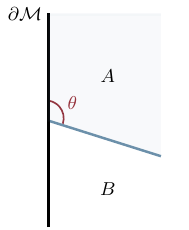}
}
\hspace{5pt}
\vspace{-5pt}
\caption{Spatial partitions of a $(2+1)$--dimensional space $\M$ with boundary $\partial\M$ (black line). (a) The region $A$ is an infinite wedge which presents a bulk corner. (b) The region $A$ is an infinite wedge adjacent to the boundary of the space, and presents a boundary corner.}
\lb{fig2}
\end{figure}
\medskip
\paragraph{Bulk corners}
\smallskip
\noindent The first partitioning of the space is the simplest one. The region $A$ is an infinite wedge with interior angle $\theta$, see Fig.\,\hyperref[fig2]{\ref{fig2}(a)}, and thus presents a corner.
Let $a(\theta)$ be the bulk corner function. It only depends on $\theta$, and by purity of the groundstate, 
\be
a(\theta)=a(2\pi-\theta)\,,
\ee
which allows us to study this corner function for \mbox{$0< \theta \le \pi$}. The bulk corner function $a(\theta)$ has other interesting properties. It is a positive convex function of $\theta$ that is decreasing on $]0,\pi]$ \cite{Hirata:2006jx}, i.e.,
\be
a(\theta)\ge0\,,\qquad \partial_\theta a(\theta)\le 0\,, \qquad  \partial^2_\theta a(\theta)\ge 0\,,\;
\ee
for $0<\theta\le\pi$. The behavior of $a(\theta)$ is constrained in the limiting regimes where the bulk corner becomes smooth $(\theta\simeq\pi)$, and where it becomes a cusp $(\theta\rightarrow0)$:
\be
a(\theta\simeq\pi) &=&\sigma\cdot(\theta-\pi)^2\,, \qquad a(\theta\rightarrow0) = \frac{\kappa}{\theta}\,,\quad\lb{alim}
\ee
where we have introduced two positive coefficients, $\sigma$ and $\kappa$.
Furthermore, the smooth bulk corner coefficient $\sigma$ is universal in the strong sense for general $3d$ CFTs, 
\be
\sigma = \frac{\pi^2}{24}C_T\,, \lb{bcuniv}
\ee  
where $C_T$ is a local observable: the central charge appearing in the two-point function of the stress tensor. This universal relation was conjectured
in \cite{Bueno:2015rda,Bueno:2015xda} and subsequently proven in \cite{Faulkner:2015csl} for general CFTs. Gapless QFTs that
  are scale and rotationally invariant, but not necessarily conformal, will also receive such a nearly-smooth corner contribution to the entanglement
  entropy. In that case, $C_T$ is replaced by a positive coefficient that appears in the so-called entanglement susceptibility \cite{WWK19}.  

\medskip
\paragraph{Corners adjacent to the boundary}
\smallskip
\noindent When the space has a boundary $\DM$, one can consider a wedge adjacent to $\DM$. In other words, the entangling surface intersects $\DM$ with an angle $\theta$, see Fig.\,\hyperref[fig2]{\ref{fig2}(b)}, defining what we call a boundary corner.
Then let $b(\theta)$ be the boundary corner function. Depending on the context, we sometimes write $b^{(\mathcal B)}(\theta)$ making the
boundary condition explicit.
The boundary corner function depends on the interior angle $\theta$ and on the boundary conditions imposed on $\DM$. By purity of the vacuum state
\be
b(\theta)=b(\pi-\theta)\,,\lb{sym}
\ee
allowing us to only consider $0<\theta\le\pi/2$. Unlike its bulk counter-part, $b(\theta)$ can be either convex or concave depending on the field theory and the boundary conditions. Its form is also constrained in the orthogonal ($\theta\simeq\pi/2$) and cusp limits:
\be
b(\theta\simeq\pi/2) &=& \eta^{\mathcal B} +\sigma^{\mathcal B}\cdot(\pi/2-\theta)^2\,,\lb{bortho}\\ 
b(\theta\rightarrow0) &=& \frac{\kappa^{\mathcal B}}{\theta}\,.\quad\lb{bcusp}
\ee

\pagebreak
At exact orthogonality, it was argued that
\begin{align}
  b(\pi/2)=\eta^{\mathcal B}\propto \cca
     \end{align}
is proportional \cite{Fursaev:2016inw,Berthiere:2016ott} to the
boundary charge $\cca$ (sometimes called $b$ in the literature) that
appears in the conformal anomaly in $3d$. Although not written explicitly here, $\cca$ does depend on the boundary condition $\mathcal B$.
We refer the reader to Appendix~\ref{apdxA} for further details regarding how the anomaly manifests itself in the trace of the stress tensor in the presence of a boundary.
Interestingly, $\cca$ was recently proved to be an RG monotone for boundary RG flows under which the bulk remains critical.
However, the coefficient $\eta^\B$ is not universal in the strong sense as its value differs for free scalars (\mbox{$\eta^{\B}=\cca/24$})
and for holographic bCFTs\footnote{Whenever holographic bCFTs are mentioned in the present paper, it refers to Takayanagi's model \cite{Takayanagi:2011zk}, see Section \ref{secHolog}.} ($\eta^{\B}=\cca /96$). Indeed, for holographic bCFTs \cite{FarajiAstaneh:2017hqv,Seminara:2017hhh}, $\eta^{\mathcal B}$ comes entirely from the anomaly, whereas for free scalars it is not the case due to the occurrence of the non-minimal coupling of the scalar field to the curvature \cite{Fursaev:2016inw}.
In Table~\ref{tab-coeff}, we summarize our findings for the coefficients appearing in the boundary corner function in the orthogonal and cusp limits for various CFTs, and the $z=2$ Lifshitz scalar.
\begin{table}[h]\renewcommand{\arraystretch}{1.5}
\begin{center}
\vspace{0.2cm}
\begin{tabular}{|c||c|c|c|c|}
\hline
 \,Theory\, & $\cca^\B$ & $\eta^\B$ & $\sigma^\B$ & $\kappa^\B$ \\ 
\hline \hline
Scalar D & \,$\displaystyle 1$\, & $1/24$ & $3/128$ & $\,0.044(4)$\\
\hline
Scalar N & $-1$ & $\;-1/24\;$ & $-1/128$ & $\;-0.024(5)\;$ \\
\hline
Dirac M & $0$ & $0$ & $1/64$ & $0.0180$ \\
\hline
$\;z=2$ Scalar D$\;$ & $\;\,$NA$\;\,$ & $1/8$ & $2/(3\pi^2)$ & $\pi/24$ \\
\hline
$z=2$ Scalar N & NA & $-1/8$ & $\;\,-1/(3\pi^2)\;$ & $-\pi/48$ \\
\hline
bEMI & $0$ & $0$ & $s_04/3$ & $s_0\pi/2$ \\
\hline
\end{tabular}
\vspace{-10pt}
\end{center}
\caption{Boundary corner coefficients in the orthogonal and cusp regimes for different critical theories. `D/N' stands for Dirichlet/Neumann, while `M' for mixed.}
\label{tab-coeff}
\end{table}

\medskip
In this manuscript, we are mostly interested in the logarithmic corner functions that appear in the entanglement entropy for regions as pictured in Fig.\,\hyperref[fig3]{\ref{fig3}}. Then according to \eqref{dual}, bulk and boundary corner functions should be related to each other through
\be
a(2\theta) &=& b^{(\B)}(\theta) + b^{(\B')}(\theta)\,,\lb{dualC}
\ee
for some boundary conditions $\B$ and $\B'$ depending on the field theory under consideration.
In what follows, we explore the implications of relations \eqref{dual} and \eqref{dualC} for various models.
\begin{figure}[h]
\centering
\includegraphics[scale=0.98]{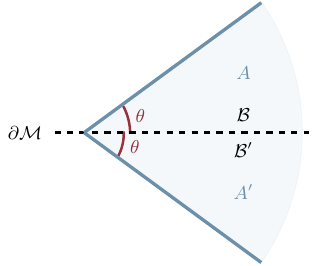}
\caption{(b)CFT$_3$ on the (half-) plane. The region $A$ and its mirror image $A'$ through $\DM$ each present a boundary corner of opening angle $\theta$, with boundary condition $\B$ and $\B'$ respectively. Their union forms a bulk corner with opening angle $2\theta$.}
\lb{fig3}
\end{figure}

\pagebreak
 
\subsection{Holographic theories}\lb{secHolog}
Within the AdS/CFT framework, certain holographic CFTs are described by a gravity theory coupled to a negative cosmological constant in one dimension higher. The holographic entanglement entropy (HEE) of some region $A$ in the boundary CFT is computed using the Ryu-Takayanagi prescription \cite{Ryu:2006bv} as the area (divided by $4G$, where $G$ is the gravitational constant) of the minimal co-dimension 2 surface homologous to $A$ on the conformal boundary of the AdS spacetime. The holographic bulk corner function $a_E(\theta)$ for $3d$ CFTs dual to Einstein gravity in AdS$_4$ has been computed in \cite{Hirata:2006jx,Myers:2012vs}.
The holographic picture of AdS/bCFT was introduced in \cite{Takayanagi:2011zk} and can briefly be sketched as follows. The dual of a bCFT$_{d}$ is given by a gravity theory in asymptotically AdS$_{d+1}$ spacetime restricted by a $d$--dimensional brane $\mathcal{Q}$ whose boundary coincides with the boundary $\DM$ of the bCFT$_{d}$. The HEE is also computed according to Ryu-Takayanagi prescription.
For the simplest geometrical setup in which the boundary of the bCFT$_3$ is flat and its extension $\mathcal{Q}$ into the bulk is completely determined by its slope $\alpha$, the HEE of an infinite wedge adjacent to the boundary was computed in \cite{Seminara:2017hhh}.
The corresponding boundary corner function $b_E^{(\alpha)}(\theta)$ depends on the extra parameter $\alpha$, which from a mathematical point of view controls the slope of the brane $\mathcal{Q}$ in the bulk, but from a field theory perspective should be related to the boundary conditions of the underlying holographic theory.

Interestingly, for the value \mbox{$\alpha=\pi/2$}, it has been observed in \cite{Seminara:2017hhh} that $b_E^{(\pi/2)}(\theta)$ is related to the holographic bulk corner function $a_E(\theta)$ as
\be
a_E(2\theta) = 2b_E^{(\pi/2)}(\theta)\,.\lb{abE}
\ee
This equality satisfies our conjecture \eqref{dual}, with boundary conditions given by $\B=\B':\a=\pi/2$. This is the unique set of values of $\a$ that leads to the relation \eqref{dual}. 

\pagebreak
Also shown in \cite{Seminara:2017hhh} was that the orthogonal-limit boundary coefficient $\sigma_E^{(\a)}$ is related to the boundary central charge $A_T^{(\a)}$ in the near-boundary expansion of the stress tensor,
\be
\sigma_E^{(\a)}=-\pi A_T^{(\a)}\,, \lb{Hratio}
\ee
where the general definition of $A_T$ in a bCFT$_d$ is \cite{Deutsch:1978sc}
\begin{align}
  \langle T_{ij}\rangle = \frac{A_T^{(\mathcal B)}}{\epsilon^{d-1}}\, \hat k_{ij}\,, \quad \epsilon \to 0\,.  
\end{align}
In the above, the stress tensor is inserted at a distance $\epsilon$ from the boundary, where we have imposed boundary condition $\mathcal B$. 
$\hat k_{ij}$ is the traceless part of the extrinsic curvature tensor of the boundary, $k_{ij}$.
The relation \eqref{Hratio} is valid for any value of the continuous parameter $\alpha$ which encodes the BCs in the holographic bCFT. A natural question to ask is whether \eqref{Hratio} holds for other theories. We address this question in Section~\ref{sec:charges}. 

\subsection{Free CFTs}

Let us first consider a non-interacting conformal scalar field with Lagrangian
density $\mathcal L=\tfrac12 \partial_\mu\phi\partial^\mu \phi$. Conformal invariance restricts the possible admissible boundary conditions to either Dirichlet or (generalized) Neumann\footnote{Generalized Neumann BC, also called Robin BC, is the generalization of Neumann BC to the case where the boundary has non-vanishing extrinsic curvature.} BCs. Then, for free scalars we conjecture that the bulk corner function $a_s(\theta)$ and the boundary corner function $b_s(\theta)$ are related through
\be
a_s(2\theta) = b_s^{(D)}(\theta) +  b_s^{(N)}(\theta)\,,\lb{corners}
\ee
where $N(D)$ stands for Neumann(Dirichlet) BCs.   

For free Dirac fermions, we consider mixed (M) BCs \cite{Luckock1991} which yield a vanishing current through the boundary, and where
a Dirichlet BC is imposed on a half of the spinor components and a Neumann BC on the other half. With these BCs, the Dirac fermion presents some similarities with scalars evenly split between Neumann and Dirichlet BCs: for example same structures of certain two-point functions \cite{McAvity:1993ue,Herzog:2017xha}, also the central charges for the Dirac fermion in the $3d$ anomaly (see \eqref{ano3}) match the sum of those for Neumann + Dirichlet scalars. We then conjecture the following relation between the bulk corner function $a_f(\theta)$ and the boundary corner function $b_f(\theta)$ for free Dirac fermions: 
\be
a_f(2\theta) = 2b_f^{(M)}(\theta)\,.\lb{fermions}
\ee
This is a special case of \eqref{dualC} with $\mathcal B=\mathcal B'=M$, similar to that for holographic bCFTs, see \eqref{abE}.
Observe that \eqref{corners} and \eqref{fermions} satisfy the reflection symmetry expected for pure states for $\theta\rightarrow\pi-\theta$. 
Using \eqref{alim} and \eqref{bortho}, in the limit $\theta\simeq\pi/2$, from \eqref{corners} and \eqref{fermions} we obtain the following relations between the bulk and boundary corner coefficients $\sigma$'s:
\be
4\sigma_s = \sigma_s^D+\sigma_s^N\,,\qquad 2\sigma_f =\sigma_f^{M}\,.\lb{sigmas}
\ee 
We can use the so-called smooth-limit boson-fermion duality \cite{Bueno:2015rda,Bueno:2015qya} $\sigma_f=2\sigma_s$ to get $\sigma_f^{M}=\sigma_s^D+\sigma_s^N$. 
One can view this last relation as a new boson-fermion duality in the presence of a boundary, which can be understood heuristically 
by recalling that a Dirac fermion with mixed BCs has two components, one with Dirichlet BCs and the other one with Neumann BCs.
In the opposite regime $\theta\rightarrow0$, inserting \eqref{alim} and \eqref{bcusp} in \eqref{corners} and \eqref{fermions} yields
\be
\kappa_s =2(\kappa_s^D+\kappa_s^N)\,, \qquad \kappa_f = 4\kappa_f^{M}\,.\lb{Kappas}
\ee

Not much is known about $b(\theta)$ for free fields, beyond $\theta=\pi/2$. Only recently \cite{Berthiere:2018ouo} has it been computed numerically on the lattice for free scalars with Dirichlet boundary conditions. Numerical values for the two boundary corner coefficients $\sigma^{\mathcal B}$ and $\kappa^{\mathcal B}$ were found to be $\sigma_s^{D}=0.023(4)\simeq3/128$ and $\kappa_s^{D}=0.044(4)$.
Then, combining this numerical result for $\sigma_s^{D}$ with \eqref{sigmas} and the well-known values of the bulk corner smooth-limit coefficients \cite{Casini:2009sr,Bueno:2015rda} $\sigma_s=1/256$ and $\sigma_f=1/128$, one can predict the boundary corner orthogonal coefficients to be
\be
\sigma_s^D\simeq\frac{3}{128}\,,\qquad \sigma_s^N\simeq-\frac{1}{128}\,,\qquad \sigma_f^M=\frac{1}{64}\,.\quad\lb{val}
\ee
For the cusp corner coefficients we have \cite{Casini:2009sr} \mbox{$\kappa_s=0.0397$} and $\kappa_f=0.0722$, which together with $\kappa_s^{D}=0.044(4)$ and \eqref{Kappas} yield
\be
\kappa_s^D=0.044(4),\;\;\, \kappa_s^N=-0.024(5),\;\;\, \kappa_f^M=0.0180.\quad\;\; \lb{valk}
\ee

Further, combining the lattice results of \cite{Berthiere:2018ouo} for Dirichlet scalars for $b_s^{(D)}$ and the exact result of \cite{Casini:2006hu,Casini:2009sr} for $a_s$, we have plotted in
Fig.\,\hyperref[fig6]{\ref{fig6}} the boundary corner function $b_s^{(N)}$ for Neumann scalars. This function is concave and negative, with a maximum at $\theta=\pi/2$. In the same figure, the boundary corner function for fermions $b_f^{(M)}$ appears, inferred from \eqref{fermions} using the results of \cite{Casini:2008as,Casini:2009sr} for the bulk corner function $a_f$. Once the functions $b(\theta)$ are properly normalized, as in Fig.\,\hyperref[fig8]{\ref{fig8}}, the corresponding curves for free scalars evenly split between Dirichlet and Neumann BCs and for free fermions with mixed BCs are very close to each other, as their bulk cousins $a(\theta)$. It will be very interesting to confront these results with direct analytical or numerical calculations of $b_s^{(N)}$ and $b_f^{(M)}$.
The numerical lattice calculation of $b_s^{(N)}$ is presented in Section~\ref{sec:lattice}; we find that the relation \eqref{corners} is indeed obeyed, thus also implying the validity of the values for the boundary coefficients for scalars with Neumann BCs predicted in \eqref{val} and \eqref{valk}.

\pagebreak
\subsubsection{Free scalars in the (half-) disk} \lb{halfdisk}
\begin{figure}[h]
\centering
\includegraphics[scale=1]{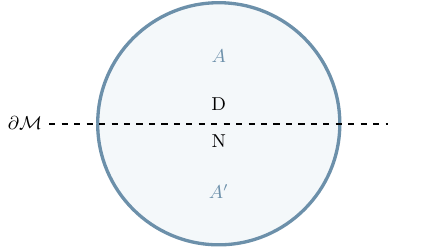}
\vspace{-3pt}
\caption{(b)CFT$_3$ on the (half-) plane. The region $A$ and its mirror image $A'$ through $\DM$, shown in light blue, are half-disks orthogonally anchored to $\partial\M$. Their union forms a complete disk.}
\lb{fig4}
\end{figure}

The Hamiltonian of a free massless real scalar field $\varphi$ in $2+1$ dimensions reads
\be
H &=& \frac{1}{2}\int d^2x\,\Big(\pi^2 + (\nabla\varphi)^2 \Big)\,.
\ee 
We consider a circular region such that we may impose either Dirichlet or Neumann BCs on its diameter, see Fig.\,\hyperref[fig4]{\ref{fig4}}. In polar coordinates $(r,\theta)$, the boundary conditions are imposed at $\theta=0,\pi$. 
Due to the symmetries, the fields can be conveniently decomposed in angular modes as
\be
\varphi(r,\theta) &=& \frac{1}{\sqrt{r}}\sum_{k}f_k(\theta)\,\varphi_{k}(r)\,,\\
\pi(r,\theta) &=& \frac{1}{\sqrt{r}}\sum_{k}f_k(\theta)\,\pi_{k}(r)\,,
\ee
where $f_k(\theta)$ is a set of orthonormal functions which depend on the BCs such that
\be
 f^{(D)}_k(\theta)&=&\sqrt{\frac{2}{\pi}}\sin(k\theta), \quad k=1,2,\cdots\,,\\
 f^{(N)}_k(\theta)&=&\sqrt{\frac{2}{\pi}}\cos(k\theta), \quad k=0,1,\cdots\,,
\ee
with $D\,(N)$ standing for Dirichlet (Neumann) BCs.
The Hamiltonian can then be written as $H = \sum_{k}H_{k}$, where
\be
H_{k} =  \frac{1}{2}\int dr \left(\pi^2_{k} +\,r\partial_r\Big(\frac{\varphi_{k}}{\sqrt{r}}\Big)^2 +\frac{k^2}{r^2}\varphi^2_{k} \right).\;\;
\lb{Hlm2}
\ee
The entanglement entropies for the half-disk with Dirichlet and Neumann BCs are thus given by
\be
S^{(D)}_{{\rm h-disk}}(R)= \sum_{k=1}^{\infty}S_{k}\,,\quad\; S^{(N)}_{{\rm h-disk}}(R)= \sum_{k=0}^{\infty}S_{k}\,,\quad
\ee
where $S_k$ is the entropy for the $k^{th}$ mode associated to $H_k$. Notice that the difference between the entanglement entropy for Dirichlet and Neumann BCs is the presence of the zero mode in the latter,
\be
S^{(N)}_{{\rm h-disk}}(R)= S_0(R) + S^{(D)}_{{\rm h-disk}}(R)\,.\lb{SdiskDN}
\ee
It is worth mentioning that the zero mode in $S^{(N)}_{{\rm h-disk}}$ contributes a factor of $1/6$ in the logarithmic part of the entropy, while the infinite sum over the higher modes, i.e. $S^{(D)}_{{\rm h-disk}}$, contributes negatively with $-1/12$.

Now, we want to compute the entanglement entropy of a complete disk of radius $R$ (no boundary here). Just as before, we can take advantage of the rotational symmetry and decompose the fields on angular modes, with eigenfunctions $f_k(\theta)=\frac{1}{\sqrt{2\pi}}e^{i k \theta}$, where $k\in\mathbb{Z}$. One then finds that the entanglement entropy of a disk is given by  
\be
S_{\rm disk}(R)&=& \sum_{k=-\infty}^{\infty}S_{k}=S_0 + 2\sum_{k=1}^{\infty}S_{k}\,.\quad\lb{Sdisk}
\ee
Comparing \eqref{Sdisk} to \eqref{SdiskDN}, one obtains
\be
S_{\rm disk}(R)= S^{(D)}_{{\rm h-disk}}(R) + S^{(N)}_{{\rm h-disk}}(R)\,,\lb{dualdisk}
\ee
which is exactly our conjectured relation \eqref{dual}, applied to the (half-) circle for the scalar field with Dirichlet/Neumann BCs. Let us emphasize that \eqref{dualdisk} is valid for the full entropies, including the finite terms. These finite contributions, let us denote them $-F_{D/N}$, are unphysical by themselves as they may be spoiled by the logarithmic term upon rescaling the UV regulator. Their sum, however, is a physical quantity $F_D+F_N=F$, that is the free energy on $\mathbb{S}^3$, see \eqref{Fdisk}.

One can also check that \eqref{dualdisk} yields a consistent relation for the corner functions: 
\be
a_s(\pi) &=& b_s^{(D)}(\pi/2) + b_s^{(N)}(\pi/2)\,\\
\Leftrightarrow\qquad 0&=& \frac{1}{24} + \frac{-1}{24}=0\,.\nonumber
\ee
Similar calculations can be done for a scalar field in a cylinder in $4d$ (see Appendix~\ref{apdxB}) or in the ($d-2$)--sphere, see e.g.\ \cite{Dowker:2010yj,Dowker:2010bu}.

\subsubsection{Lattice calculations for the free scalar} \label{sec:lattice} 
We consider the discretized Hamiltonian of a 2+1 dimensional free massless scalar field on a square lattice given by
\be
H &=& \frac{1}{2}\sum_{x,y}\Big[ \pi^2_{x,y} + (\phi_{x+1,y}-\phi_{x,y})^2 
 + (\phi_{x,y+1}-\phi_{x,y})^2 \Big],\nonumber\\
\lb{Hd}
\ee
where \mbox{$\mathbf{x} = (x,y)$} represents the spatial lattice coordinates with $x_i=1, \cdots,L_i$, and $L_i$ is the lattice length along the $i^{th}$ direction. The total number of sites is $N=L_xL_y$.
The Hamiltonian \eqref{Hd} corresponds to a lattice of coupled quantum harmonic oscillators, and its linearly dispersing acoustic mode
is described by the free scalar CFT. 
$H$ may also be written more compactly as
\be
H = \frac{1}{2}\sum_{\mathbf{x}} \pi^2_{\mathbf{x}} +  \frac{1}{2}\sum_{\,\mathbf{x},\mathbf{x}'}\phi_{\mathbf{x}}K_{\mathbf{x}\mathbf{x}'}\phi_{\mathbf{x}'} ,\lb{Hd2}
\ee
where $K$ is an $N\times N$ matrix encoding the nearest-neighbor interactions between lattice sites as well as the boundary conditions. 
The vacuum two-point correlation functions \mbox{$X_{\mathbf{x}\mathbf{x}'}\equiv\la \phi_\mathbf{x}\phi_{\mathbf{x}'}\ra$} and $P_{\mathbf{x}\mathbf{x}'}\equiv\la \pi_\mathbf{x}\pi_{\mathbf{x}'}\ra$ are given in terms of the matrix $K$ by
\be
X = \frac{1}{2}K^{-1/2}\,, \quad{\rm and}\quad P  = \frac{1}{2}K^{1/2}\,. \lb{corr}
\ee
The entanglement entropy can then be calculated \cite{Casini:2009sr} from the eigenvalues $\nu_\ell$ of the matrix $C_A=\sqrt{X_AP_A}$, where $X_A$ and $P_A$ are the correlation matrices restricted to the region $A$:
\be
S(A) &=& \sum_{\ell} \Bigg[\Big(\nu_\ell+\frac{1}{2}\Big)\log\Big(\nu_\ell+\frac{1}{2}\Big) \nonumber\\
&& \hspace{1.5cm} -\Big(\nu_\ell-\frac{1}{2}\Big)\log\Big(\nu_\ell-\frac{1}{2}\Big) \Bigg].\quad\lb{EE}
\ee

We choose to impose periodic BC in the $x$ direction and Dirichlet-Neumann BCs in the $y$ direction, i.e. \mbox{$\phi_{L_x+1,y}=\phi_{1,y}$}, and $\phi_{x,0}=0$ and \mbox{$\phi_{x,L_y+1}-\phi_{x,L_y}=0$.} Note that the Dirichlet-Neumann BCs do not have the zero mode that would have been present for Neumann-Neumann.
We compute the entanglement entropy for regions $A$ of width $L_x^A$ with fixed ratio $L_x^A/L_y=4$, as depicted in \mbox{Fig.\,\hyperref[fig5]{\ref{fig5}}}, and extract the logarithmic contribution by performing least-squares fits of our numerical data to the scaling ansatz \cite{Casini:2006hu,Casini:2009sr,Helmes:2016fcp,DeNobili:2016nmj}
\be
S(L_y)&=& s_1 L _y- 2s_{\rm log}\log L_y + s_0 \lb{fit}\\ 
&&\quad + s_{-1}L_y^{-1} + \cdots + s_{-p_{\rm max}}L_y^{-p_{\rm max}}\,.\nonumber
\ee
For the Dirichlet-Neumann BCs that we have chosen, the region $A$ displays four boundary corners; two Dirichlet and two Neumann (the factor two is due to the symmetry $b(\theta)=b(\pi-\theta)$). The logarithmic contribution $2s_{\rm log}$ in the entropy is thus the sum of the Dirichlet and Neumann boundary corners functions, such that once extracted, we may directly check our conjectured relation \eqref{corners} as
\be
s_{\rm log} = b^{(D)}(\theta) +  b^{(N)}(\theta) \stackrel{?}{=} a(2\theta)\,.
\ee 
\begin{figure}[h]
\centering\vspace{-5pt}
\includegraphics[]{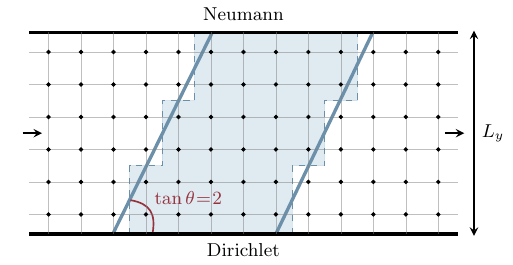}\vspace{-5pt}
\caption{Two-dimensional square lattice with Dirichlet-Neumann BCs imposed in the (vertical) $y$ direction and PBCs in the (horizontal) $x$ direction. 
The region $A$ is shown in blue, and has $L_x^A=5$. The entangling surface intersects the boundaries with angles $\theta=\arctan(\pm 2)$.}
\lb{fig5}
\end{figure}
\begin{figure}
\centering\hspace*{-6pt}
\includegraphics[scale=1.04]{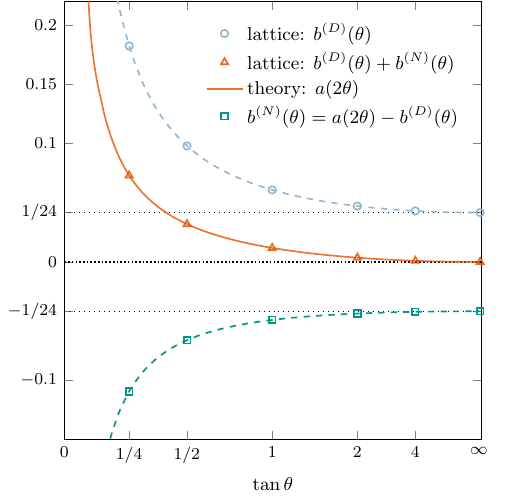}\vspace{-5pt}
\caption{Corner entanglement for free scalars. The orange triangles are our numerical results for $b^{(D)}(\theta)+b^{(N)}(\theta)$, while the solid orange line is the ``high precision ansatz'' of \cite{Helmes:2016fcp} for $a(2\theta)$. The numerical results for $b^{(D)}(\theta)$ found in \cite{Berthiere:2018ouo} are the blue circles. With green squares are shown the values of the Neumann boundary corner function as deduced 
from $b^{(N)}(\theta)=a(2\theta)-b^{(D)}(\theta)$. Finally, the dashed blue and green lines are interpolations of the numerical data.}
\lb{fig6}
\vspace{-7pt}
\end{figure} 

We present in Appendix~\ref{apdxC} the implementations of different boundary conditions on a one-dimensional lattice, and in particular Neumann BC. The extension to higher dimensional lattices is straightforward.
The two-dimensional vacuum two-point functions in the thermodynamic limit $L_x\rightarrow\infty$ are the following:
\be
\la \phi_{i,j} \phi_{r,s}\ra  &=&  \frac{\binom{i-r-1/2}{i-r}}{L_y+1/2}\sum_{k_y}\sin(k_y j)\sin(k_y s)\nonumber\\
&& \hspace{-10pt}\times\sqrt{\frac{z^{2(i-r)+1}}{1-z^2}} \,{_2}F_1\bigg(\frac{1}{2},\frac{1}{2};i-r+1; \frac{z^2}{z^2-1}\bigg)\,,\nonumber\\
&&\lb{Xang}\\
\la \pi_{i,j}\pi_{r,s}\ra &=&  \frac{\binom{i-r-3/2}{i-r}}{L_y+1/2}\sum_{k_y}\sin(k_y j)\sin(k_y s)\nonumber\\
&& \hspace{-10pt}\times \sqrt{\frac{1-z^2}{z^{2(i-r)+1}}}\,{_2}F_1\bigg(\hspace{-4pt}-\frac{1}{2},\frac{3}{2};i-r+1; \frac{z^2}{z^2-1}\bigg)\,,\nonumber\\
&&\lb{Pang}
\ee
where we defined \mbox{$z = \Big(|\sin(k_y/2)| - \sqrt{\sin^2(k_y/2)+1}\Big)^2$} with $k_y=\pi (2n_y-1)/(2L_y+1)$ and $n_y=1,\,\cdots,L_y$.
Expressions \eqref{Xang} and \eqref{Pang} are the matrix elements of the correlation matrices $X_A$ and $P_A$ respectively (where $(i, j)$ and $(r, s)$ are the raw and column indices respectively). 
On square lattices, angles which obey $\tan\theta = r\in \mathbb{Q}$ are accessible by ``pixelation" of the region $A$ (see e.g. \cite{Helmes:2016fcp,Berthiere:2018ouo}). This is shown in \mbox{Fig.\,\hyperref[fig5]{\ref{fig5}}} for $\tan\theta=\pm2$. Our lattice results for the free scalar with Dirichlet-Neumann BCs are given in Table~\ref{tab1} in which we have reported the digits that we found to be robust. We also include in this table the values of $a(2\theta)$ from the ``high precision ansatz'' of \cite{Helmes:2016fcp} (see Appendix~\ref{apdxD}), the numerical results of \cite{Berthiere:2018ouo} for $b^{(D)}(\theta)$, as well as the values of $b^{(N)}(\theta)$ deduced from the previous results. Plots of all this are shown in Fig.\,\hyperref[fig6]{\ref{fig6}}.

As can be seen in Table~\ref{tab1}, we find a difference of less than $0.5\%$ between our numerical results for $b^{(D)}(\theta)+b^{(N)}(\theta)$ and the field theoretic
ones \cite{Helmes:2016fcp} for $a(2\theta)$, thus implying the validity of \eqref{corners}. The high precision lattice results \cite{Helmes:2016fcp} for the bulk corner function $a(2\theta)$ are also in close agreement with our numerical results; we do not show their values here since they agree with the field theoretic ones within error bars.
Further, we have computed the $n=2$ R\'enyi entropy and find that \eqref{corners} also holds in that case within less than $1\%$ discrepancy between the numerics and the theory. Table~\ref{tab1} shows the comparison with the high precision field theory results for $a_2(2\theta)$ \cite{Helmes:2016fcp}.   

Using our numerical results, we find that the R\'enyi index $n$ and the angle dependences in the entropy do not factorize. If it were the case, we would have \mbox{$b_n(\theta)/b(\theta)=const$} valid for all angles $\theta$. This ratio for $n=2$ shows a deviation of $13\%$ for Neumann BCs, and
only $2\%$ for Dirichlet BCs between $\theta=\pi/2$ and $\theta=\arctan(1/4)$.
At orthogonality, our results for $n=1,\,2$ are in perfect agreement with the following relation \cite{Fursaev:2016inw,Berthiere:2016ott}
\be
b_n(\pi/2)=\frac{1}{2}\left(1+\frac{1}n\right)b(\pi/2)\,, \lb{factorize}
\ee
where $b(\theta)\equiv b_1(\theta)$. This can be understood by using the replica trick. The R\'enyi entropies may be computed by introducing in the underlying manifold a conical singularity located at the entangling surface. In three dimensions, when a flat entangling curve intersects orthogonally the flat physical boundary, the singular spacetime factorizes as the product of a two-dimensional cone (the singular part, $n$--dependent) with a semi-infinite interval (the entangling line). As a result, the R\'enyi entropy is simply proportional to the entanglement entropy, hence \eqref{factorize}. Now if the entangling curve is not orthogonal to the boundary, we do not have a product space, therefore the R\'enyi index $n$ and the angle dependences in the entropy do not factorize, as we verified numerically.
\begin{widetext}
\begin{table*}[h]\renewcommand{\arraystretch}{1.5}  
\begin{center}
\begin{tabular}{|c|c|c|c|c|c|c|c|c|}
\hline
$n$ & \multicolumn{4}{|l}{\hspace{1.5cm} Entanglement entropy $n=1$} & \multicolumn{4}{|l|}{\hspace{2cm} R\'enyi entropy $n=2$}\\
\hline
 \,$\tan\theta$\, & $\,b^{(D)}(\theta)+b^{(N)}(\theta)$ & \;$a(2\theta)$ \cite{Helmes:2016fcp} & \;$b^{(D)}(\theta)$ \cite{Berthiere:2018ouo} & \;\;$b^{(N)}(\theta)$\;\; & $\,b_2^{(D)}(\theta)+b_2^{(N)}(\theta)$ & \;$a_2(2\theta)$ \cite{Helmes:2016fcp} & \;\,$b_2^{(D)}(\theta)$\; & \;\;$b_2^{(N)}(\theta)$\;\;  \\ 
\hline \hline
1/4 & $0.0730$ & \;$0.0730(6)$\; & $0.182(4)$ & $\,-0.109(4)\,$ & $0.0412$ & $0.04127$ & $0.1340$ & $-0.0927$  \\
\hline
1/2 & $0.0319$ & $0.03195$ & $0.09798$ & $-0.0660$ & $0.0177$ & $0.01779$ & \;$0.07223$\; & $-0.05444$ \\
\hline
1 & $0.0118(3)$ & $0.011833$ & $0.06081$ & $-0.04898$ & $0.00648$ & $0.006487$ & $0.04511$ & $-0.03862$ \\
\hline
2 & $0.00357$ & $0.003579$ & $0.04717$ & $-0.04359$ & $0.00194$ & $0.001943$ & $0.03522$ & $-0.03327$ \\
\hline
4 & $\;0.00095\;$ & $0.000953$ & $0.04310$ & $-0.04215$ & $0.00051$ & $0.000516$ & $\;0.03228\;$ & $-0.03177$  \\
\hline
$\infty$ & $10^{-7}$ & $0$ & \;$\sim1/24$\; & \,$\sim-1/24$\, & $10^{-8}$ & $0$ & $\sim 1/32$ & $\;\sim -1/32\;$  \\
\hline
\end{tabular}
\vspace{-8pt}
\end{center}
\caption{Lattice results for the boundary corner entanglement for free scalars. The second column presents our numerical results for $b^{(D)}(\theta)+b^{(N)}(\theta)$, which we compare to those of \cite{Helmes:2016fcp} for $a(2\theta)$ in the third column. In the fourth column are reported the numerical results of $b^{(D)}(\theta)$ \cite{Berthiere:2018ouo}. Next, we give values of the Neumann boundary corner function $b^{(N)}(\theta)=a(2\theta)-b^{(D)}(\theta)$.
The next 2 columns compare our numerical results for the $n=2$ R\'enyi case $b_2^{(D)}(\theta)+b_2^{(N)}(\theta)$ with the theoretical one $a_2(2\theta)$ of \cite{Helmes:2016fcp}. We also give $b_2^{(D)}(\theta)$, which was computed using a lattice with DD boundary conditions. The last column shows $b_2^{(N)}(\theta)=a_2(2\theta)-b_2^{(D)}(\theta)$. }
\label{tab1}
\end{table*}\vspace{-10pt}
\end{widetext}

\subsection{Relation to central charges} \lb{sec:charges}
It has been conjectured in \cite{Berthiere:2018ouo} that relation \eqref{Hratio} should hold for free scalars split evenly between Neumann and Dirichlet BCs, and for free fermions with mixed BCs, due to properties that these theories share with the holographic one at $\alpha=\pi/2$. 
For scalars, the value of $A_T^s$ for both BCs is known, $A_T^{s,D}=A_T^{s,N}=-1/(128\pi)$, which is actually independent of the boundary condition.
The expression corresponding to \eqref{Hratio} for Dirichlet $+$ Neumann scalars reads
\be
(\sigma_s^D+\sigma_s^N)/2=-\pi A^s_T\,,\lb{Hscal}
\ee
and is indeed satisfied with the values of $\sigma_s^{D/N}$ given in \eqref{val}. Note that \eqref{Hratio} does not hold for free scalars with Dirichlet \textit{or} Neumann BCs alone \cite{Berthiere:2018ouo}. The value of $A_T$ for fermions is known through its relation with the boundary central charge $c$ in the trace anomaly \cite{Miao:2017aba} (see Appendix~\ref{apdxA}), \mbox{$A^f_T=-1/(64\pi)=2A_T^{s}$} hence 
\be
\sigma_f^M=-\pi A^f_T\lb{Hferm}
\ee
holds for fermions as well, using $\sigma_f^M=1/64$ predicted in \eqref{val}. As we will see shortly, this may be understood as a consequence of $A_T$ being related to $C_T$ for free scalars and fermions. The validity of $\sigma^\B=-\pi A_T$ for free Dirichlet-Neumann scalars, mixed fermions and holographic theories dual to Einstein gravity raises the question whether it also holds for other $3d$ theories with appropriate BCs. It would be interesting to test this hypothesis with different models in order to see if universality is indeed at play here.

Now, recall that for bulk corners, the smooth-limit coefficient $\sigma$ is universal and proportional to $C_T$, see \eqref{bcuniv}. Then, using \eqref{bcuniv} and \eqref{sigmas} together with \eqref{Hscal} and \eqref{Hferm} yields the relation
\be
A_T = -\frac{\pi}{12}C_T\,. \lb{AC}
\ee
One can check that this equality indeed holds for scalars with $A^s_T=-1/(128\pi)$ and $C^s_T=3/(32\pi^2)$, and for fermions using $A^f_T=-1/(64\pi)$ and $C^f_T=3/(16\pi^2)$. We thus find through the connection between bulk and boundary corner entanglement that the charge $A_T$ appearing in the near-boundary expansion of the stress tensor is in fact related to $C_T$, and it appears so in a universal way for free fields.
In fact, such a relation between $A_T$ and $C_T$ seems to exist in any dimensions for free fields, and for holographic theories with BC $\a=\pi/2$ only \cite{Miao:2018qkc}, see Appendix~\ref{apdxA} for further details.

We also notice that with \eqref{val}, the boundary corner coefficients for free fields may be expressed in a universal form
\be
\sigma^{\mathcal B }=\frac{\pi^2}{12}C_T+\frac{\cca}{64} = -\pi A_T+\frac{\cca}{64}\,,\lb{sigbuniv}
\ee
where $\cca$ is the boundary central charge in the conformal anomaly (see Appendix~\ref{apdxA}): $\cca=\pm 1$ for scalars with Dirichlet $(+)$ and Neumann $(-)$ BCs, and $\cca=0$ for fermions with mixed BCs. 
Note that \eqref{sigbuniv} is not valid for holographic bCFTs with arbitrary $\a$, but it does hold for $\a=\pi/2$ (the charge $\cca\propto\cot\a$ vanishes in that case).

\section{Extensive Mutual Information model} \label{sec:bEMI}

Within the Extensive Mutual Information model (EMI) \cite{Casini:2005rm,Casini:2008wt,Swingle:2010jz}, the entanglement entropy of a region $A$ in infinite flat space is obtained by the following double integral over two copies of the boundary $\partial A$ of $A$:
\be
S_\emi(A)=s_0\int_{\partial A} d\mathbf{r'}\int_{\partial A} d\mathbf{r} \,\frac{\hat{n}\cdot\hat{n}'}{|\mathbf{r'}-\mathbf{r}|^{2(d-2)}}\,, \lb{SEMI}
\ee
where $d$ is the spacetime dimension, $s_0$ is a positive constant, and $\hat{n}$ is an outward pointing vector normal to $\partial A$.
The EMI model has the interesting property that the mutual information, \mbox{$I(A, B) = S(A) + S(B) - S(A \cup B)$}, satisfies the extensivity property:
\be
I(A,B \cup C) = I(A,B) + I(A,C)\,,
\ee
hence its name.\pagebreak

The entanglement entropy given by \eqref{SEMI} is valid in flat space without boundaries.
We introduce the following generalization that includes a flat boundary $\DM$ by the following simple ansatz, which we dub `bEMI':
\be
S_\bemi(A) = \frac{1}{2} S_\emi(A\cup A')\,, \lb{SEMIB}
\ee
where $A'$ is the mirror image of $A$ with respect to $\DM$, see Fig.\,\hyperref[fig7]{\ref{fig7}}. By construction, $S_\bemi$ satisfies \eqref{dual} with identical boundary conditions
$\mathcal B=\mathcal B'$, although we are being agnostic about the physical meaning of the boundary condition since we do not know what theory
  has an entanglement entropy given by the bEMI. Note that we refer to \eqref{SEMI} and \eqref{SEMIB} as entanglement entropies, but keep in mind that the EMI and bEMI ansatzes can be extended to general R\'enyi entropies by replacing $s_0$ with $s_{0,n}$.

\subsection{Corner entanglement in $2+1$ dimensions}
For the EMI model, the bulk corner function $a_\emi(\theta)$ reads \cite{Casini:2008wt}
\be
a_\emi(\theta) = 2s_0\big(1+(\pi-\theta)\cot\theta\big)\,. \lb{a_emi}
\ee 

Our bEMI ansatz thus yields the boundary corner function $b_\emi(\theta)$:
\be
b_\emi(\theta)&=&\frac{1}{2}a_\emi(2\theta)\,.\lb{bemi}
\ee
We note that this relation is identical to that of the free Dirac fermion \eqref{fermions} with mixed BCs,
scalars with mixed BCs, and to the holographic one \eqref{abE} with BCs $\a=\pi/2$. 
Using \eqref{a_emi}, we find that the boundary corner function vanishes at orthogonality $b_\emi(\pi/2)=0$,
which implies the vanishing of the central charge
  \begin{align}
   \cca^{\rm bEMI} =0\,.
  \end{align}
  The expansion coefficients for angles near $\pi/2$ and $0$ read $\sigma^{\rm bEMI}=s_0 4/3$ and $\kappa^{\rm bEMI}=s_0\pi/2$, respectively.
  These coefficients are listed in Table~\ref{tab-coeff}. 
Using the known value \cite{Bueno:2015rda} for the bulk theory,
$C_T=s_016/\pi^2$, we see that the following relation holds: 
\begin{align}
  \sigma^{\rm bEMI} = \frac{\pi^2}{12}\, C_T \,,
\end{align}
which is also satisfied by a free Dirac fermion with mixed BCs, free scalars with Dirichlet-Neumann BCs, and holographic CFTs with $\alpha=\pi/2$. 
Now, assuming the relation $\sigma^{(\mathcal B)}=-\pi A_T^{(\mathcal B)}$ holds for the bEMI, we can extract the boundary central charge:
$A_T^{\rm bEMI}=- s_0 4/(3\pi)$. We note that this value is the same as the one we would have obtained using $A_T^{\rm bEMI}=-\pi C_T/12$.
However, since we do not know whether these relations hold for the bEMI, the value of $A_T$ is a conjecture.

$b_\emi(\theta)$ (normalized) is plotted as a function of $\theta$ in Fig.\,\hyperref[fig8]{\ref{fig8}}. As one may see in this figure, the normalized boundary corner functions for the bEMI, holography, fermions, and N$+$D scalars are hardly discernible. Universality seems to be at play here. Gaining a better understanding of this is of foremost importance.

\subsection{$(1+1)-$dimensional systems}
In $d=2$, the two integrals in \eqref{SEMI} should be replaced by a double sum over the set of endpoints $p_i$ of the intervals for which the entropy is computed:
\be
S_\emi(A_1\cup A_2\cup\cdots)&=&-s_0\sum_{i,j}\hat{n}_i\cdot\hat{n}_j \log|p_i-p_j|\,.\nonumber
\ee\vspace{-40pt}
\be
\,\lb{EMI2}
\ee
At coincidental points $p_i=p_j$, the expression above needs to be regulated; we thus introduce a short-distance UV cut-off $\eps$, i.e. $|p_i-p_i|\rightarrow\eps$.
Let us denote the set of endpoints by \mbox{$\{p_i\}\equiv\{u_i,v_i\}$}, where $u_i$ and $v_i$ are the left and right endpoints of the interval $A_i$, respectively. In the basis $(0,\hat{e}_x)$ with the unit vector $\hat{e}_x$ in the direction of increasing $x$, the normal vectors $\hat{n}_i$ at $p_i$ are simply $\hat{n}_i=\pm\hat{e}_x$, depending on the endpoint being left ($-$) or right ($+$). It is then straightforward to show that the $p$-intervals entropy for the EMI model in $1+1$ dimensions takes the form:
\be
S_\emi(A_1\cup A_2\cup\cdots\cup A_p)&=&\lb{SEMIbn}\\
&&\hspace{-3.cm}2s_0\log\bigg(\frac{\prod_{i,j}|u_i-v_j|}{\eps^p\prod_{i<j}|u_i-u_j||v_i-v_j|}\bigg)\,.\nonumber\hspace{0.3cm}
\ee
Setting $s_0=\frac{1}{12}(1+\frac{1}{n})$, the entropy \eqref{SEMIbn} is exactly the $n$--R\'enyi entropy of a free massless Dirac fermion \cite{Casini:2005rm}!

Our bEMI ansatz \eqref{SEMIB} for $p$ regions yields
\be
S_\bemi(A_1\cup\cdots\cup A_p)&=&\lb{bSEMIbn}\\
&&\hspace{-2.8cm}\frac{1}{2}S_\emi(A'_p\cup \cdots\cup A'_1\cup A_1\cup\cdots\cup A_p)\,.\quad\nonumber
\ee
In particular, for one interval of length $\ell$ connected to the boundary in $1+1$ dimensions, eq.\,\eqref{bSEMIbn} gives
\be
S_\bemi(\ell) = \frac{1}{12}\left(1+\frac{1}{n}\right)\log\frac{2\ell}{\eps}\,,
\ee
which is exactly the result \eqref{bEE2} for a Dirac fermion (with Virasoro central charge $c=1$) on the semi-infinite line. For one interval of length $\ell$ at a distance $d$ from the boundary, we obtain
\be
S_\bemi(\ell,d) &=& \frac{1}{12}\left(1+\frac{1}{n}\right)\log\frac{4\ell^2d(\ell+d)}{\eps^2(\ell+2d)^2}\,,\;\;
\ee
which again perfectly agrees with the known result \cite{Fagotti:2010cc} for the free fermion in a semi-infinite system. Note that taking the limits $d\rightarrow\infty$ and $d\rightarrow0$, one recovers \eqref{EE2} and \eqref{bEE2}, respectively.
We therefore conclude that the EMI and bEMI models are exact for free fermions in $1+1$ dimensions.

\begin{figure}[h]
\centering
\includegraphics[scale=1]{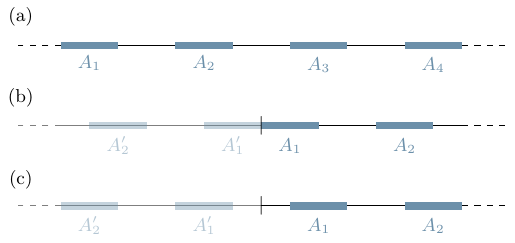}\vspace{-3pt}
\caption{Multi-interval entanglement for the (b)EMI model. (a) Four intervals on the infinite line without boundary. (b) Two intervals on the semi-infinite line, with $A_1$ connected to the boundary, and the mirror image through the boundary on the left. (c) Two intervals on the semi-infinite line, none connected to the boundary, and the mirror image through the boundary on the left.}
\lb{fig7}
\vspace{-5pt}
\end{figure}

\pagebreak
\section{Lifshitz field theory in $2+1$ dimensions} \label{sec:lif}
Lifshitz field theories (LFTs) are non-relativistic theories which exhibit anisotropic scaling between space and time, with characteristic dynamical exponent $z\neq 1$. In 2+1 dimensions, the free Lifshitz real scalar theory with dynamical critical exponent $z = 2$ enjoys many interesting features. The corresponding Euclidean action for the non-compact scalar $\varphi$ in $d=3$ is
\be
I_{\rm LFT}[\varphi]=\frac12 \int d^3 x \left[ (\partial_\tau\varphi)^2 + (\nabla^2\varphi)^2 \right].
\ee
We have absorbed an inessential constant that would appear in front of the term with spatial derivatives by using field and coordinate rescalings.
For this model, the groundstate wavefunctional is given in terms of the Euclidean action $I_{\rm CFT}[\varphi]$
of the \emph{two-dimensional} CFT \cite{Ardonne:2003wa},
\be
\ket{\Psi} = \frac{1}{\sqrt{Z}}\int [d\varphi] e^{-\frac{1}{2}I_{\rm CFT}[\varphi]}\ket{\varphi}\,,\lb{GS}
\ee
where $Z$ is the partition function of the CFT,
\be
Z = \int [d\varphi] e^{-I_{\rm CFT}[\varphi]}\,, \quad I_{\rm CFT}[\varphi] = \frac{1}{2}\int d^2x \,(\nabla\varphi)^2 .\;\;\quad
\ee
The groundstate wavefunction \eqref{GS} of the $z=2$ free scalar is thus conformally
invariant in space \cite{Ardonne:2003wa}!

We consider spatial bipartitions such as those shown in Fig.\,\hyperref[fig2]{\ref{fig2}}. Then, provided $\varphi$ is non-compact, the R\'enyi entanglement entropy for the groundstate is given by \cite{Fradkin:2006mb}
\be
S_n=-\log\frac{Z_A Z_B}{Z_{A\cup B}}\,,
\ee
and is independent of the R\'enyi index $n$, which we henceforth drop. $Z_{A}$ and $Z_{B}$ are the free (CFT) scalar partition functions on regions $A$ and $B$, respectively, with continuity of the fields requiring Dirichlet BCs on the entangling curve $\Sigma$. $Z_{A\cup B}$ is the partition function on the entire space $\M$, with specified boundary conditions, e.g. Dirichlet or Neumann BCs, on the space boundary $\DM$. In [\onlinecite{Fradkin:2006mb}], only Dirichlet BCs were considered.
The entanglement entropy can thus be written as the difference in free energies 
\be
S=F_A+F_B-F_{A\cup B}\,.\lb{Sz2}
\ee
For the free scalar field, the free energy can be expressed in terms of the heat kernel
$K(s)\equiv e^{s\triangle}$ of the ($2d$ in our case) Laplacian operator $\triangle$,
\be
F = -\frac{1}{2} \int_{\eps^2}^\infty \frac{ds}{s}\, \tr\, K(s)\,,\lb{FK}
\ee
where the trace is taken over the region of interest and $\eps\to 0$.
Computing the entanglement entropy \eqref{Sz2} thus boils down to computing the trace of the heat kernel on the three domains $A$, $B$, and $A\cup B$.

\subsection{Corner entanglement for the $z=2$ scalar}

Suppose a two-dimensional domain $\M$ has a piecewise smooth boundary $\cup_i \DM_i$ consisting of a number of $\DM_i$ (with extrinsic curvature $k_i$) which may intersect at some points, the corners. Either Dirichlet or Neumann BC is imposed on each of the pieces $\DM_i^{D/N}$, thus yielding three types of corners (NN, DD and ND). The heat trace $\tr\, K(s)$ admits an asymptotic expansion as $s\rightarrow 0$ of the form\footnote{This classical asymptotic expansion of the heat trace may break down at the $p=3$ level when considering the $N/D$ problem (e.g. corners with mixed BCs), see \cite{Dowker:2000vm}. However, we are only interested in the heat coefficients up to $p=2$.}:
\be
\tr\, K(s) \simeq \sum_{p\ge0} \text{a}_p\, s^{(p-2)/2}\,,\lb{trK}
\ee
where the coefficients $\text{a}_p$ depend on the geometry of the domain and on the boundary conditions. Plugging the heat trace expansion in \eqref{FK} yields the following leading terms in the free energy:
\be
F=-\frac{{\rm a}_0}{2\eps^2}-\frac{{\rm a}_1}{\eps} - {\rm a}_2\log\frac{\ell}{\eps} + \O(1), 
\ee
where $\ell$ is a length scale characteristic of the size of the domain on which the free energy is computed.
The first three heat coefficients are given by (see \cite{Vassilevich:2003xt} and references therein)
\be
{\rm a}_0 &=& \frac{1}{4\pi}\int_\M 1 \,,\\
{\rm a}_1 &=&  \frac{1}{8\sqrt{\pi}}\Big(\sum_i\int_{\DM_i^N}1 - \sum_j\int_{\DM_j^D}1\Big)\,,\\
{\rm a}_2 &=&  \frac{1}{24\pi}\Big(\int_\M R + \sum_i \int_{\DM_i} 2k_i\Big) \nonumber \\
&& \qquad + \sum_j f_H(\theta_j)+ \sum_k f_M(\theta_k)  \,,
\ee
where we have defined the following heat corner functions
\be	
f_H(\theta) &=& \frac{1}{24}\left(\frac{\pi}{\theta}-\frac{\theta}{\pi}\right),\\
f_M(\theta) &=& -\frac{1}{48}\left(\frac{\pi}{\theta}+\frac{2\theta}{\pi}\right), \lb{MC}
\ee
where H stands for a corner with homogenous BCs (DD or NN) and M for a mixed corner (ND). Note that the mixed heat corner coefficient can be obtained by applying relation \eqref{dualC}, that is 
\be
f_H(2\theta) =f_H(\theta)  + f_M(\theta)\,.\lb{hcorners}
\ee
One can explicitly check \eqref{MC} by computing the heat trace on mixed wedges of opening angles, e.g., $\pi/2,\,\pi/4,\,\pi/6$ with the method of images. This result for the mixed corner was previously obtained with the same arguments by Dowker in \cite{Dowker:2000bi}.
Notice that $f_M(\theta)$ is not a monotonic function of $\theta$ over $[0,\pi]$ as $f_H(\theta)$.

Getting back on track, it is clear that the volume terms, i.e.\;the a$_0$'s, do not contribute to the entropy, while the boundary terms a$_1$ produce the area law (due to the Dirichlet BC imposed on the entangling surface). The first two (smooth) terms in a$_2$ do not contribute to the entropy either. However, the last two terms in a$_2$, originating from the corners, give rise to a logarithmic scaling in the entropy. The corner functions corresponding to the geometries in Fig.\,\hyperref[fig2]{\ref{fig2}} are easily obtained by summing the heat coefficients $f_{H/M}$ for the regions $A$ and $B$ and subtracting those for $A\cup B$. The entanglement entropy for the $z=2$ free scalar field thus has the following form:
\be
S &=&B \frac{\ell}{\eps} - s_{\rm log}\log\frac{\ell}{\eps} + \O(1) \,,\;\lb{EEz2}
\ee
where the logarithmic coefficient $s_{\rm log}$ is given by the different corner functions,
\be
s_{\rm log} &=&\sum_{i} a_\LFT(\theta_i) + \sum_{j} b_\LFT(\theta_j) \,.
\ee
Below we give formulas for these corner functions which will allow us to explicitly check our conjecture \eqref{dual} for this theory.
\subsubsection{Bulk corner}

The well-known \cite{Fradkin:2006mb} bulk corner function $a_\LFT(\theta)$ for the wedge does not depend on the boundary conditions on $\DM$, and reads for the $z=2$ free scalar:  
\be
a_\LFT(\theta) &=& \frac{(\pi-\theta)^2}{12\theta(2\pi-\theta)} \,,\lb{atheta}
\ee
which implies that the smooth- and cusp-limit coefficients respectively read \cite{corner-bounds15} 
  \begin{align} 
    \sigma=1/(12\pi^2)\, ,\quad \kappa = \pi/24\,.
  \end{align}

\subsubsection{Boundary corner}

The boundary corner function $b_\LFT(\theta)$ depends on the boundary condition imposed on $\DM$ (either D or N),
\be
b_\LFT^{(D)}(\theta) &=& \frac{1}{24}\left(\frac{\pi-\theta}{\theta}+\frac{\pi}{\pi-\theta}\right), \label{bLifD}\\
b_\LFT^{(N)}(\theta) &=& -\frac{1}{48}\left(\frac{\pi+2\theta}{\theta}+\frac{\pi}{\pi-\theta}\right), \label{bLifN}
\ee
with
\begin{align}
  b_\LFT^{(D)}(\pi/2) &=-b_\LFT^{(N)}(\pi/2)=1/8\,, \\
  \sigma^{D}&=-2 \sigma^{N}=2/(3\pi^2)\,, \\
  \kappa^{D}&=-2 \kappa^{N}=\pi/24\,.
\end{align}
These coefficients are listed in Table~\ref{tab-coeff}.  
The two functions $b_\LFT^{(D,N)}$ display the same qualitative behaviors as their relativistic cousins. Indeed, $b_\LFT^{(D)}$ is a positive convex function of $\theta$ as $b_s^{(D)}$, while $b_\LFT^{(N)}$ is negative and concave as $b_s^{(N)}$, as may be seen in Fig.\,\hyperref[fig8]{\ref{fig8}}. 
Surprisingly, the normalized functions $b_s^{(D)}$ and $b_\LFT^{(D)}$ plotted in Fig.\,\hyperref[fig8]{\ref{fig8}} coincide almost perfectly. 
This is unexpected given how different the two theories are (relativistic-conformal versus 
non-relativistic). However, such an agreement does not occur for Neumann BC.

Remarkably, the corner functions for the $z=2$ free scalar satisfy the same conjectured equality \eqref{corners} as for the free relativistic scalar field,
\be
a_\LFT(2\theta) = b_\LFT^{(D)}(\theta) + b_\LFT^{(N)}(\theta)\,.\lb{dualz2}
\ee
This exact result gives us further confidence in the validity of \eqref{dual} and \eqref{dualC} for certain QFTs.

\begin{figure}[h]
\centering\hspace*{-7pt}
\includegraphics[scale=1]{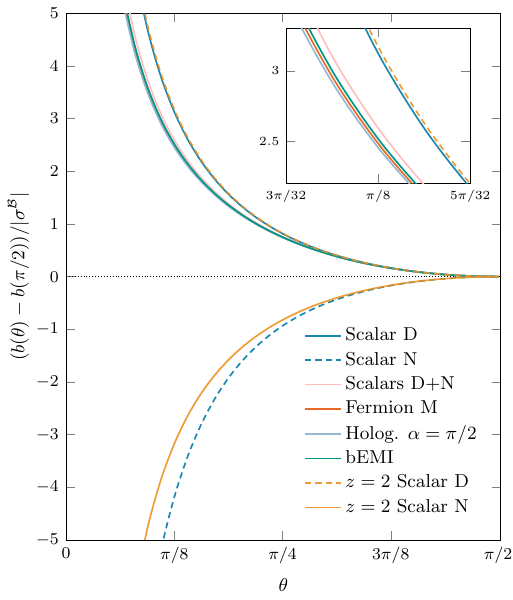}
\vspace{-6pt}
\caption{Boundary corner entanglement for various theories. The boundary corner functions are normalized in such a way that near $\theta=\pi/2$ they behave as $b(\theta\simeq\pi/2)=\pm(\pi/2-\theta)^2$. The inset shows a closeup of the positive curves.} 
\lb{fig8}
\end{figure}

\section{Massive theories} 
\label{sec:massive}
So far, we have only considered gapless theories. However, many QFTs are not gapless, and so it is highly desirable to understand the fate of our bulk-boundary relation in that case. For one, we expect our relation \eqref{dual} between bulk and boundary entropies to hold for certain free massive theories. As an example, let us take the free massive scalar field. The arguments presented in Section~\ref{hkernel} should carry through to the massive case. Indeed, the heat kernel for a massive scalar field is simply obtained from the massless case as $K^{(m)} = e^{-m^2s}K^{(m=0)}$, such that $K^{(m)}=K^{(m)}_N + K^{(m)}_D$ holds. One could also repeat the treatment for the (half) disk geometry in 
Section~\ref{halfdisk} for the massive case. The Hamiltonian \eqref{Hlm2} with a mass term is obtained by replacing $k^2\rightarrow k^2+m^2$, thus relation \eqref{dualdisk} also holds for free massive scalars.

In the half-space, for a flat entangling surface that intersect orthogonally the physical boundary, the corresponding entanglement entropy can be computed explicitly in any dimensions \cite{Berthiere:2016ott}. For instance, in $3d$ one has 
\be
S_{D/N}^{(m)}=B \frac{\ell}{\eps} \pm \frac{1}{24}\log(\eps m) 
+ \cdots\,,
\ee
which satisfies the bulk-boundary relation \eqref{dual}: the logarithms cancel when we add the entropies corresponding to the two boundary conditions.
The ellipsis represents terms subleading in $\epsilon$; $\ell$ is the IR cut-off for the size of the entangling region.

The case of massive Dirac fermions must be treated with care because gapless edge states can be present on the boundary, which would arise in the description
of Chern or $Z_2$ topological insulators, for instance. These gapless edge modes can affect the 
  entanglement entropy of regions touching the boundary. We leave the discussion of such effects for future work.

\section{Conclusion} \label{sec:conclusion}

We studied the quantum entanglement properties of systems in the presence of a physical boundary.
We have proposed a bulk-boundary relation \eqref{dual} relating the R\'enyi entropies of certain theories with  and without a boundary.  
Our attention was focused on situations where the entangling surface intersects the boundary of the space. In particular, in three dimensions, this leads to a new type of corner, called a boundary corner, from which originates new kinds of universal quantities in the entanglement entropy. These corner-induced logarithmic terms are not to be confused with those arising in the bulk when the entangling surface presents a singularity. For a given theory, the corresponding boundary corner function $b^{(\B)}(\theta)$ depends on the opening angle $\theta$ of the corner adjacent to the physical boundary and on the boundary conditions $\B$. Our bulk-boundary relation connects the universal bulk and boundary corner terms for a family of theories \eqref{dualC}. The relation applies for boundary theories with ``mixed'' BCs, such that for bCFTs the Euler boundary central charge vanishes $\cca=0$, see \eqref{anomE}. This is the case for free scalars evenly split between Dirichlet and Neumann BCs, as well as free Dirac fermions
  with mixed BCs, holographic CFTs with an $\alpha=\pi/2$ BC, and the boundary Extensive Mutual Information Model (bEMI).
  The latter allows a simple \mbox{geometric} calculation of the entanglement entropy in the presence of a flat boundary, and thus constitutes a very useful tool.  

We also studied the Lifshitz free scalar with dynamical exponent $z=2$.
The bulk and boundary corner functions can be computed explicitly, producing remarkably simple functions of the opening angle $\theta$
for both Dirichlet and Neumann BCs. These functions satisfy the bulk-boundary relation \eqref{dualC}, and behave very similarly to the case of the
relativistic scalar. In particular, the Neumann corner function \eqref{bLifN} is negative for all angles, just as in the relativistic case.
 
An interesting direction would be to study the relation between the bulk and boundary entanglement entropies of other Lifshitz theories and CFTs, such as the Ising CFT or its $N>1$ cousins (generally
  known as $O(N)$ Wilson-Fisher fixed points). In tour-de-force numerical calculations, the bulk corner function $a(\theta)$ for angles of $\pi/2$ was studied 
  on the lattice \cite{PhysRevLett.110.135702,pitch,Kallin:2014oka,Sahoo:2015hma}, and analytically in the large-$N$ limit \cite{Whitsitt17}. 
  It would be worthwhile to apply these methods to corners adjacent to the boundary, for different boundary conditions.

Our results also generalize to higher dimensions. For instance, we discuss the case of cylindrical entangling regions in $3+1$ dimensions in Appendix \ref{apdxB}.
  More interestingly, one could study the case of trihedral vertices, where three planes meet at a point.
  These vertices lead to a logarithmic contribution to the entanglement entropy for gapless theories, and were studied recently in the bulk for critical states \cite{Singh14,Hayward17,Bednik18,WWK19,Bueno2019}, but much remains unknown about their properties. One could examine how
the bulk trihedral entropy relates to that of boundary trihedral corners, where the two planes forming the entangling surface intersect the flat physical boundary to form a trihedral vertex.      

\begin{acknowledgments}
We would like to thank Sergue\"i Tchoumakov for useful discussions, and his assistance with the computing cluster.  We also acknowledge interesting discussions with Jacopo Sisti and Jia Tian. 
C.B. thanks the Universit\'e de Montr\'eal for warm hospitality during the completion of this project.
C.B. was supported in part by the National \mbox{Natural} Science Foundation of China (NSFC, Nos. 11335012, 11325522, 11735001), and by a Boya Postdoctoral Fellowship at Peking University.
W.W.-K.\ was funded by the Fondation Courtois, a Discovery Grant from NSERC, a Canada Research Chair, and a 
``\'Etablissement de nouveaux chercheurs et de nouvelles chercheuses universitaires'' grant from FRQNT.
This research was enabled in part by support provided by Calcul Qu\'ebec (www.calculquebec.ca)
and Compute Canada (www.computecanada.ca). 
\end{acknowledgments}

\appendix
\addtocontents{toc}{\protect\setcounter{tocdepth}{1}}

\section{Comments on bulk and boundary charges}\lb{apdxA}
Boundary conformal field theories offer a wider bestiary of central charges than conformal field theories. This has of course to be imputed to the presence of the `b' in bCFT. 
In three dimensional spacetimes with boundaries, the conformal anomaly no longer vanishes and there are two boundary charges,
$\cca$ and $c$ \cite{Graham:1999pm,Solodukhin:2015eca}. The vacuum expectation value of the trace of the stress tensor integrated over the spacetime reads
\be
\int_{\M_3}\la T_\mu^\mu \ra = -\frac{\cca}{96}\chi[\DM_3]+\frac{c}{256\pi}\int_{\DM_3}\tr\,\hat{k}^2,\quad \lb{ano3}
\ee
where $\chi[\DM_3]$ is the Euler characteristic of the boundary and $\hat{k}_{\mu\nu}$ is the traceless part of the extrinsic curvature tensor of the boundary. The charge $c$ is independent of boundary conditions for free fields, while $\cca$ for scalars is not. For a free scalar field, $c=1$ and $\cca=\pm1$ for Dirichlet ($+$) and Neumann ($-$) boundary conditions, and $c=2$, $\cca=0$ for a free Dirac fermion with mixed boundary conditions.
Recently \cite{Herzog:2017kkj,Miao:2017aba}, $c$ has been connected to two other boundary charges, namely $A_T$ and $c_{nn}$, where $c_{nn}$ is the charge in the two-point function of the displacement operator. 
Then, with eq.\,\eqref{AC} which relates $A_T$ to $C_T$ for free fields, one finds that all the boundary charges presented above, with the exception of $\cca$, are related to the bulk charge $C_T$,
\be
A_T&=&-\frac{c}{128\pi}=-\frac{\pi}{16}c_{nn}= -\frac{\pi}{12}C_T  \,.\lb{bcharges}
\ee
Therefore, only the boundary charge $\cca$ and the bulk charge $C_T$ are independent for free fields. 
One may also wonder if such a relation between $A_T$ and $C_T$ exists in higher dimensions.
For scalars, fermions and vectors we find in the literature \cite{Deutsch:1978sc,Osborn:1993cr,Perlmutter:2013gua,Miao:2017aba}
\begin{alignat}{3}
&C_T^s=\frac{d}{d-1}\frac{\Gamma^2(d/2)}{4\pi^d}\,, \;\;\quad A_T^s=-\frac{\Gamma(d/2)}{2^d\pi^{d/2}(d^2-1)} \,, \vspace{5pt}\\
& C_T^f=\frac{2^{\lfloor d/2\rfloor}d}{8\pi^d}\Gamma^2(d/2)\,, \hspace{10pt} A_T^{f(4d)}=-\frac{1}{40\pi^2}\,,\vspace{5pt}\\
&C_T^{v(4d)}=\frac{4}{\pi^4}\,, \hspace{50pt} A_T^{v(4d)}=-\frac{1}{20\pi^2}\,.
\end{alignat}
For scalars, one gets in $d$ dimensions
\be
A_T = -\frac{2^{2-d}\pi^{d/2}}{d(d+1)\Gamma(d/2)}C_T\,. \lb{univ}
\ee
One can check that \eqref{univ} is actually satisfied for every known values of $C_T$ and $A_T$ for free CFTs. As an interacting example, for holographic bCFTs we have in $d$ dimensions \cite{Seminara:2017hhh}, 
\be
A_{T,E}^{(\alpha)}&=&-\frac{L^{d-1}_{AdS}}{8\pi G}\Bigg[\frac{1}{\cos\alpha}\,_2F_1(-1/2,(2-d)/2;1/2;\cos^2\alpha)\nonumber\\
&&\qquad\qquad -\frac{\sin^{d-2}\alpha}{\cos\alpha}+\frac{\sqrt{\pi}\,\Gamma(d/2)}{\Gamma\big(\frac{d-1}{2}\big)} \Bigg]^{-1},\\
C_{T,E}&=&\frac{L^{d-1}_{AdS}}{8\pi G}\frac{(d+1)!}{(d-1)\pi^{d/2}}\frac{1}{\Gamma(d/2)}\,,
\ee
and it is easy to show that for $\alpha=\pi/2$ we have
\be
A_{T,E}^{(\pi/2)} = -\frac{2^{2-d}\pi^{d/2}}{d(d+1)\Gamma(d/2)}C_{T,E}   \,, \;\;\quad\lb{univH}
\ee
which is exactly $\eqref{univ}$.

In $d=4$ bCFTs, the conformal anomaly reads \cite{Herzog:2015ioa,Solodukhin:2015eca}
\be
\int_{{\cal M}_4}\la T\ra&=&-\frac{a}{180}\chi[{\cal M}_4]+\frac{c}{1920\pi^2}\int_{{\cal M}_4} W^2_{\mu\nu\alpha\beta} \quad\lb{ano4}\\
&&\hspace{-1.cm}-\frac{b_1}{240\pi^2}\int_{\partial{\cal M}_4} \hat{k}^{\mu\nu}W_{\mu nn\nu}+\frac{b_2}{280\pi^2}\int_{\partial{\cal M}_4}\hspace{-4pt}\Tr\hat{k}^3\,.\nonumber
\ee
The coefficients $b_1$ and $b_2$ are new boundary central charges while $a$ and $c$ are the well-known bulk charges. Only $b_2$ depends on boundary conditions as one finds from free fields $b_1=c$. The values of these charges for free fields are given by
\begin{alignat}{3}
a^s&=1\,,\qquad   &a^f&=11\,,\qquad  &a^v&=62\,, \quad\\
c^s&=1\,,\qquad   &c^f&= 6\,,\qquad  &c^v&=12\,, \\
b_2^{s,D(N)}&=1\,(7/9)\,,\qquad  &b_2^{f}&=5\,, \qquad  &b_2^v&=8\,.
\end{alignat}
It is known that $b_1=c=3\pi^4C_T$ for free fields. In \cite{Herzog:2017xha}, it was proven that $b_1$ is related to the coefficient $c_{nn}$ in the displacement operator two-point function as $b_1=2\pi^4c_{nn}$. Further, in \cite{Miao:2017aba} $b_1$ has been connected to $A_T$ via $b_1=-240\pi^2A_T$. Thus through this chain of relations for $b_1$, we have for free fields
\be
A_T&=&-\frac{\pi^2}{120}c_{nn}=-\frac{b_1}{240\pi^2}=-\frac{\pi^2}{80}C_T \,.
\ee
The last equality involving $A_T$ and $C_T$ is exactly relation \eqref{univ} for $d=4$.
Interestingly, the coefficient $c_{nn}$ in $d$ dimensions has been related to $A_T$ in \cite{Miao:2018dvm},
\be
A_T= -\frac{d\,\pi^{(d-1)/2}}{d-1}\frac{\Gamma\big(\frac{d-1}{2}\big)}{\Gamma(d+2)}\,c_{nn}\,,
\ee
where $\displaystyle c_{nn}^s= \frac{\Gamma^2(d/2)}{2\pi^d}$ for free scalars \cite{McAvity:1993ue}. Then one can use  \eqref{univ} to predict the value $\displaystyle c_{nn}^f=\frac{d-1}{4\pi^d}2^{\lfloor d/2\rfloor}\Gamma^2(d/2)$ for fermions in any dimensions.
This last expression agrees with the known values for fermions in $d=3,\,4$ dimensions.

\section{Cylinders in $d=4$ dimensions}\lb{apdxB}
\begin{figure}[h]
\begin{center}
\includegraphics[scale=0.9]{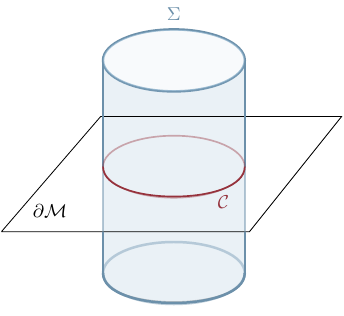}
\end{center}
\vspace{-0.4cm}
\caption{(3+1)--dimensional spacetime (time slice). The entangling surface $\Sigma$ is a two-dimensional cylinder  orthogonal to the boundary $\DM$. Their intersection (red) is a circle.}
\label{fig9}
\end{figure}

Let us consider the entanglement entropy of a scalar field in a cylinder of length $L/2$ and radius $R\ll L$, anchored orthogonally on the flat boundary of the space $\DM$, as depicted in Fig.\,\hyperref[fig9]{\ref{fig9}}.
We use cylindrical coordinates $(r,\theta,z)$. We may impose either Dirichlet or Neumann BCs on the boundaries at $z=0,\,L/2$. 
In a similar manner as for the disk, we can dimensionally reduce our problem from $3+1$ to $1+1$ dimensions. To that end, the fields are decomposed in angular and axial modes as
\be
\varphi(r,\theta,z) &=& \frac{1}{\sqrt{r}}\sum_{kl}f_{kl}(\theta,z)\,\varphi_{kl}(r)\,,\\
\pi(r,\theta,z) &=& \frac{1}{\sqrt{r}}\sum_{kl}f_{kl}(\theta,z)\,\pi_{kl}(r)\,,
\ee
where $f_{kl}(\theta, z)$ is a set of orthonormal functions which depend on the BCs such that
\be
 f^{(D)}_{kl}(\theta,z)&=&\frac{1}{\sqrt{\pi L}}e^{il\theta}\sin(2\pi k z/L), \quad\hspace{1pt} k=1,2,\cdots\,,\qquad\;\;\\
 f^{(N)}_{kl}(\theta,z)&=&\frac{}{\sqrt{\pi L}}e^{il\theta}\cos(2\pi k z/L), \quad k=0,1,\cdots\,,
\ee
and $l\in\mathbb{Z}$.
The Hamiltonian can then be written as $H = \sum_{kl}H_{kl}$, where
\be
H_{kl} &=&  \frac{1}{2}\int dr \left(\pi^2_{kl} +\,r\partial_r\Big(\frac{\varphi_{kl}}{\sqrt{r}}\Big)^2 +\Big(\frac{l^2}{r^2}+\omega_k^2\Big)\varphi^2_{kl} \right),\;\nonumber\\
\lb{Hln2}
\ee
and $\omega_k=2\pi k/L$.
The entanglement entropies for Dirichlet and Neumann BCs are thus given by
\be
S^{(D)}_{\rm cyl}(R)&=&  \sum_{l=-\infty}^{\infty}\sum_{k=1}^{\infty}S_{kl}\,,\\
S^{(N)}_{\rm cyl}(R)&=& \sum_{l=-\infty}^{\infty}\sum_{k=0}^{\infty}S_{kl}\,,
\ee
where $S_{kl}$ is the entropy associated to $H_{kl}$.

\nopagebreak
Now, we want to compute the entanglement entropy of a cylinder of length $L$ and radius $R\ll L$ (no boundary here). It is convenient to compactify the \mbox{direction $z$} by imposing periodic BCs $z=z+L$ and decompose the fields on axial and angular modes with eigenfunctions \mbox{$f_{kl}(\theta,z)=\frac{1}{\sqrt{2\pi L}}e^{i l \theta}e^{i 2\pi k z/L}$}, where $k,l \in\mathbb{Z}$. The entanglement entropy of a cylinder is thus given by 
\be
S_{\rm cyl}(R)= \sum_{k,l=-\infty}^{\infty}S_{kl}\,,
\ee
which can be written as
\be
S_{\rm cyl}(R)= S^{(D)}_{\rm cyl}(R) + S^{(N)}_{\rm cyl}(R)\lb{cyltot} \,,
\ee
as for the disk case.

Again, it is interesting that the difference between the entanglement entropy for Dirichlet and Neumann BCs is the presence of the $k=0$ mode in the latter. One further notices that the entropy associated to this mode is in fact the entropy of a scalar field in a disk of radius $R$ in $2+1$ dimensions \eqref{Sdisk}, and we have
\be
S^{(N)}_{\rm cyl}(R)=  S^{(D)}_{\rm cyl}(R) +S_{\rm disk}(R) \,.\lb{ScylDN}
\ee
The equality \eqref{ScylDN} yields the following relation for the logarithmic contributions $s_{\rm cyl}$:
\be
s^{(N)}_{\rm cyl} = s^{(D)}_{\rm cyl}\,,\lb{cylND}
\ee
as there is no logarithmic contribution for the disk in $2+1$ dimensions. Equation \eqref{cylND} is actually the expected result for the cylinder. In a flat four-dimensional spacetime with a flat boundary $\DM$, the logarithmic term in the entanglement entropy for an entangling surface $\Sigma$ intersecting orthogonally the boundary is given by \cite{Solodukhin:2008dh,Fursaev:2013mxa}
\be
s_{\rm log}\; =\; \frac{a}{180}\chi[\Sigma] + \frac{c}{240\pi}\int_\Sigma \tr\, \hat{k}_i^2\,.
\ee
The first term is the Euler characteristic of $\Sigma$ and $(\hat{k}_i)_{\mu\nu}$ is the traceless part of the extrinsic curvature of $\Sigma$ as embedded in the four-dimensional spacetime. The central charges $a$ and $c$ do not depend on the BCs. The Euler characteristic of a cylinder (with a geodesic boundary or none) is zero and only the $c$-part in the logarithmic contribution remains.

A similar calculation for an hemisphere would yield the same relations as \eqref{cyltot} and \eqref{cylND}, only for the (hemi)sphere, it is the $a$-part that is non-vanishing. Note that $\chi[{\rm sphere}]=2$ and $\chi[{\rm hemisphere}]=1$.

\section{Implementation of boundary conditions for the discretized scalar field}\lb{apdxC}

 The continuum Hamiltonian of a free massless scalar field in $1+1$ spacetime dimensions is
 \be
H= \frac12 \int dx \left(\pi^2 + \phi (-\partial^2_x)\phi \right)\,.
 \ee
In the discrete case, the fields are evaluated at a lattice site $i\in[1,N]$ such that $\phi(x)\rightarrow\phi(x_i)\equiv\phi_i$ and \mbox{$\pi(x)\rightarrow\pi(x_i)\equiv\pi_i$}. The above Hamiltonian is thus replaced by 
 \be
 H= \frac12 \left(\pi^T\pi + \phi^T K\phi \right)\,, \lb{Hdiscrete}
 \ee
 where $\phi^T=(\phi_1 ,\phi_2,\cdots, \phi_N)$, $\pi^T=(\pi_1 ,\pi_2, \cdots, \pi_N)$, and the matrix $K$ is the discretized version of the spatial laplacian operator $-\partial^2_x$. In the static case, the Hamiltonian \eqref{Hdiscrete} yields the equations of motion 
\be
  K\phi=0\,,\lb{eomsD}
\ee 
with specified boundary conditions at both ends of the lattice. Since we are considering a scalar field, its discrete counter-part is the harmonic chain with nearest neighbors interactions. 
The equation of motion for the oscillator $\phi_i$ reads:
\be
-\phi_{i-1} + 2\phi_i -\phi_{i+1} = 0\,.\quad\lb{eoms}
\ee
One should however take the boundary conditions into account in the equations of motion of $\phi_1$ and $\phi_N$. In order to implement boundary conditions on a discrete domain, we first introduce fictitious degrees of freedom, $\phi_0$ and $\phi_{N+1}$. The equations of motion for $\phi_1$ and $\phi_N$ are
\be
-\phi_{0} + 2\phi_1 -\phi_{2} &=& 0\,,\quad\lb{eom1}\\
-\phi_{N-1} + 2\phi_N -\phi_{N+1} &=& 0\,,\lb{eomN}\quad
\ee
but we can get rid of the extra $\phi_0$ and $\phi_{N+1}$ by substituting in \eqref{eom1} and \eqref{eomN} boundary conditions such as
\begin{alignat}{2}
{\rm Periodic}&:\quad   &\phi_0&=\phi_{N}\,, \quad\\
{\rm Dirichlet}&:\quad   &\phi_0&=0\,,\\
{\rm Neumann}&:\quad  &\phi_1&-\phi_0=0\,,
\end{alignat}
at $i=0$, and similarly at $i=N+1$. Then, the equations of motion including the boundary conditions are put in the vector form \eqref{eomsD}, from which one can read off the matrix $K$. For example, with Dirichlet/Neumann BC on the left/right end, $K$ is a tridiagonal $N\times N$ matrix,
 \be
 K_{DN}=
 \left(\begin{array}{ccccc}
 2 & -1 &  &  &  \\
 -1 & 2 & -1 & & \\
    & \ddots & \ddots & \ddots &  \\
    &  & -1 & 2 & -1 \\
    &  &  & -1 & 1
 \end{array}\right).\quad
 \ee
 
 The matrix $K$ has eigenvectors $v_{i,j}$ ($i$ labels the components of $j$th eigenvector) and eigenvalues \mbox{$\omega^2_j=4\sin^2\left(k_j/2\right)$} where $k_j$ depends on the BCs:
 \begingroup
\allowdisplaybreaks
 \begin{alignat}{3}
{\rm P:}&\; &k_j&=\frac{2\pi j}{N}\,, \; &v_{i,j}&=\frac{1}{\sqrt{N}}\exp(\imath k_j i)\,,\quad\\
{\rm DD:}&\;   &k_j&=\frac{\pi j}{N+1}\,,\; &v_{i,j}&=\sqrt{\frac{2}{N+1}}\sin(k_j i)\,,\\
{\rm NN:}&\;  &k_j&=\frac{\pi(j-1)}{N}\,,\; &v_{i,j}&=\sqrt{\frac{2-\delta_{j,1}}{N}}\cos[k_j (i-1/2)]\,,\;\;\\
{\rm DN:}&\;  &k_j&=\frac{\pi(2j-1)}{2N+1}\,,\; &v_{i,j}&=\sqrt{\frac{2}{N+1/2}}\sin(k_j i) ,\\
{\rm ND:}&\;  &k_j&=\frac{\pi(2j-1)}{2N+1}\,,\; &v_{i,j}&=\sqrt{\frac{2}{N+1/2}}\cos[k_j (i-1/2)] \,.\;\;\,
\end{alignat}
\endgroup
Finally, we obtain the groundstate correlation functions for the scalar field on the lattice as
\be
\la \phi_i \phi_j\ra &=&  \frac{1}{2} K_{ij}^{-1/2} =  \frac{1}{2}\sum_{n}\omega_n^{-1}v_{i,n}v^\dagger_{j,n}\,,\\
\la \pi_i \pi_j\ra &=&  \frac{1}{2} K_{ij}^{1/2} =  \frac{1}{2}\sum_{n}\omega_nv_{i,n}v^\dagger_{j,n}\,.
\ee
Note that for a massive field we have $\omega^2_n\rightarrow \omega^2_n+m^2$.

\section{High precision ansatz for the scalar bulk corner function}\lb{apdxD}
We present in this appendix the high precision ansatz of \cite{Helmes:2016fcp} for the scalar bulk corner function $a_n(\theta)$, where $n$ is the R\'enyi index. This ansatz takes the form:
\be
a_n(\theta)\simeq \sum_{p=1}^M\sigma_n^{(p-1)}(\theta-\pi)^{2p} + \frac{2\kappa_n}{\pi^{2M+1}}\frac{(\theta-\pi)^{2(M+1)}}{\theta(2\pi-\theta)}\,,\nonumber\\
\ee
where $M$ corresponds to the number of smooth limit coefficients $\sigma_n^{(p-1)}$ used ($\sigma_n^{(0)}\equiv\sigma_n$).
We refer the reader to \cite{Casini:2009sr} for the details regarding the expansion of the corner function in the nearly smooth limit. 
We give in Table~\ref{tab2} the coefficients $\sigma_n^{(p-1)}$ up to $p=8$ ($M=8$) for $n=1,2$ found in \cite{Casini:2009sr,Helmes:2016fcp}. For the cusp limit coefficients, the value of $\kappa\equiv\kappa_1$ is reported below eq.\,\eqref{val}, while the $n=2$ one may be found in \cite{Bueno:2015qya}, $\kappa_2=0.0227998$.
\begin{table}[h]\renewcommand{\arraystretch}{1.5}
\begin{center}
\begin{tabular}{|c|c|c|}
\hline
 \,$n$\, & $1$ & $2$ \\ 
\hline \hline
$\sigma_n$ & \,$\displaystyle\frac{1}{256}$\,\rule[-2ex]{0pt}{5.5ex} & \,$\displaystyle\frac{1}{48\pi^2}$\, \\
\hline
$\sigma_n^{(1)}$ & \,$\displaystyle\frac{20+3\pi^2}{18432\pi^2}$\,\rule[-2ex]{0pt}{5.7ex} & \,$\displaystyle\frac{5+\pi^2}{960\pi^4}$\, \\
\hline
$\sigma_n^{(2)}\times10^5$  & \,$2.67327749$\, & \,$1.55767377$\, \\
\hline
$\sigma_n^{(3)}\times10^6$ & \,$2.70080311$\, & \,$1.56206308$\, \\
\hline
$\sigma_n^{(4)}\times10^7$ & \,$2.72879243$\, & \,$1.57369200$\, \\
\hline
$\sigma_n^{(5)}\times10^8$ & \,$2.75578382$\, & \,$1.58861117$\, \\
\hline
$\sigma_n^{(6)}\times10^9$ & \,$2.78590964$\, & \,$1.60561386$\, \\
\hline
\,$\sigma_n^{(7)}\times10^{10}$\, & \,$2.81790229$\, & \,$1.62402979$\, \\
\hline
\end{tabular}
\end{center}
\caption{Smooth limit coefficients for the scalar bulk corner function $a_n(\theta)$ \cite{Helmes:2016fcp}.}
\label{tab2}
\end{table}

\newpage

\bibliographystyle{utphys} 

\begingroup

\endgroup

\end{document}